# Why User Studies and Participant Experience Reporting Matter for VR Motion Privacy?

AZIM IBRAGIMOV, University of Florida, USA
ERIC D. RAGAN, University of Florida, USA

Public VR game leaderboards contain tracked motion recordings uploaded by hundreds of thousands of users. Once uploaded, these recordings are accessible to anyone and create privacy risks (i.e., identification and profiling). Prior work has proposed mechanisms that modify tracked movement to reduce these risks. Their utility is commonly evaluated through physical deviation, where smaller deviations indicate better utility, while user studies are less common. However, it remains unclear how well physical deviation explains users' acceptance of a mechanism compared to user studies. We examine this through a user study of three VR motion privacy mechanisms at five physical deviation levels. We find that user studies explain substantially more variation in mechanism acceptance than physical deviation, although physical deviation remains significant. We also find that prior VR experience and exposure to VR privacy mechanisms significantly affect acceptance. We recommend combining physical deviation with user studies and reporting participants' prior experience.

## 1 Introduction

Virtual reality (VR) devices commonly rely on motion tracking to support 3D interaction. Tracked head and hand movements allow users to navigate virtual environments, manipulate objects, and interact with others. Many VR games (e.g., Beat Saber [4], Tilt Brush [10], Echo VR [1], and Rumble VR [6]) allow users to record and share their gameplay, including their tracked movements. Some of these games also have public leaderboards where players compete based on their scores. For example, Beat Saber has public leaderboards where users can submit their scores. To submit a score, users are also required to submit a motion recording of their gameplay (i.e., their tracked head and hand movements). The leaderboard system uses this recording to verify the submitted score and help prevent cheating. As a result, public VR game leaderboards contain motion recordings from hundreds of thousands of users [32]. Once submitted, these recordings remain on public servers where anyone can access them [35]. Although these recordings are needed for score verification, they also create persistent records of users' movements [30].

This persistence creates privacy risks because VR motion can function as a behavioral biometric, making users recognizable from their movement patterns in ways analogous to fingerprints [34]. For example, Nair et al. [30] showed that 55,541 Beat Saber users could be re-identified from motion recordings uploaded to a public leaderboard. Beyond identification, prior studies have shown that VR motion can reveal other sensitive information about users, including demographic [15, 42, 48], health [15, 18, 47, 48], cognitive [15, 29], and behavioral [11, 15] information. Protecting persistent VR motion data has therefore become an important privacy challenge.

Researchers have proposed numerous mechanisms that protect VR motion data by modifying recorded movements to reduce how easily users can be recognized from their motion recordings [16, 31, 39, 44]. These mechanisms modify

Authors' Contact Information: Azim Ibragimov, University of Florida, Gainesville, Florida, USA; Eric D. Ragan, University of Florida, Gainesville, Florida, USA.

the recording so that the user's movements appear different from their original movements. They can do so in different ways: Gaussian noise [16, 33, 44] can make movements appear jittery; smoothing [2, 16, 44] can make movements appear smoother; and temporal downsampling [16, 27, 44] can make movements appear discontinuous. These modifications make users harder to identify, but they can also change how users' performance is represented in the protected recording: a person who normally moves quickly and makes sudden movements may appear slower and smoother after a smoothing privacy mechanism is applied. Similarly, users' movements may appear less precise, delayed, or otherwise different from how they actually performed. As a result, users may find some of these modifications less acceptable because the protected recording may no longer represent their performance in the way they expect.

Because these modifications can affect whether users find protected recordings acceptable, it is important to evaluate the utility of privacy mechanisms in addition to the privacy they provide [15, 38]. The most common way to evaluate utility is through distance-based metrics, which measure the physical deviation between the original and privatized motion [3, 16, 26, 37, 39, 40]. Under these evaluations, smaller deviations indicate better utility because the privatized motion remains closer to the user's original motion.

More recently, some studies have evaluated utility through user studies, where participants directly interact with or evaluate privatized motion [31, 43, 44]. These studies suggest that as privacy mechanisms are applied at stronger levels, which results in greater deviation from the original motion, their effects on user experience may differ across mechanisms. For example, Wilson et al. [44] found that increasing mechanism strength affected the usability of Gaussian noise more substantially than smoothing. Their results suggest that greater changes to the original motion do not necessarily affect users in the same way across different privacy mechanisms. This raises an important question for distance-based evaluations: does the amount of physical deviation introduced by a privacy mechanism correspond to how users perceive and accept the resulting motion? This relationship may also differ across users depending on their prior VR experience and exposure to VR privacy mechanisms.

We address this gap through a controlled user study of VR motion privacy mechanisms. In the study, participants viewed privatized motion recordings from a Beat Saber leaderboard. Each participant viewed five recordings that were privatized using the same randomly assigned privacy mechanism: Gaussian noise, smoothing, or temporal downsampling. The recordings were shown at five deviation levels: 1, 3, 5, 7, and 11 cm. This created a mixed-design study with *privacy mechanism* varied between participants and *deviation level* varied within participants. Participants were asked to rate the recordings based on perceived *spatial degradation* (whether movements appeared spatially inaccurate), *perceived temporal degradation* (whether movements appeared delayed or incorrectly timed), *perceived privacy risk* (whether the motion could still be used to identify or track the user), *trust* (whether participants felt confident relying on the privacy mechanism), and *acceptance* (whether participants were willing to share the protected recording). Additionally, we examined whether participants' prior experience with VR and VR privacy mechanisms had an impact on their acceptance.

This paper makes the following contributions:

(1) A controlled user study with 286 participants evaluating three VR motion privacy mechanisms across five physical deviation levels (§4).
(2) A comparison of physical deviation and user study evaluations for understanding VR privacy mechanism acceptance (§5.3). We find that both physical deviation and user study evaluations are significantly associated with privacy mechanism acceptance, but user study evaluations explain substantially more variation in acceptance than physical deviation alone. Based on these findings, we recommend combining physical deviation with user study evaluations rather than relying on either evaluation alone.

(3) An analysis of whether prior VR experience and prior exposure to VR privacy mechanisms shape sharing acceptance across mechanism types and physical deviation levels (§5.4). We find that participants' prior experience significantly affects sharing acceptance. Based on these findings, we recommend that future user studies collect and report participants' prior experience with VR and VR privacy mechanisms and consider this experience when interpreting study results.

## 2 Background and Related Work

The following section provides additional background on privacy risks associated with VR motion data (§2.1) and how VR privacy mechanisms can mitigate these risks (§2.2).

### 2.1 Privacy Risks of VR Motion Data

Motion has long been studied as a behavioral biometric, with gait recognition being one of the most well-known examples [17]. Prior studies have shown that gait can identify individuals at scale, including in populations of up to 62,528 individuals [14, 23, 24, 46]. Traditionally, however, gait recognition has relied on motion extracted from video footage captured by external cameras. As the distance between the user and camera increases, lower image resolution, occlusion, and variations in viewing conditions can make accurate motion extraction more difficult [45].

VR devices introduce a new concern because they are equipped with sensors that capture users' movements with high precision. Unlike camera-based gait recognition, these sensors are worn by the user and positioned close to the body. As a result, one of the main challenges of gait recognition, reliably extracting motion from distant video footage, is largely eliminated in VR systems [15].

Another concern is that people may perceive the same motion data differently depending on how it is collected. Studies of camera-based surveillance have found that people may view continuous recording and monitoring as intrusive, particularly when footage can be used to identify or track them without their knowledge [41]. In VR, however, users may treat motion data more casually, even though these data can reveal similarly identifying information. BeatLeader[1], a public leaderboard and replay-sharing platform for Beat Saber, illustrates this contrast. Users can upload motion recordings captured by their headsets and controllers, making them publicly accessible and downloadable by anyone. To date, more than 100,000 users have willingly uploaded their motion recordings to the platform [32].

Once an attacker accesses these recordings, several privacy risks become possible. Prior research has shown that VR motion can be used for account linking, where an attacker identifies anonymous users from their motion and links their identities across separate recordings [15, 16, 30, 44]. Thus, changing a username or avatar may not be enough to remain anonymous because users' motion can still be distinctive. In the largest study to date, Nair et al. linked over 4 million recordings from more than 55,000 users with 94% accuracy [30]. Prior studies have also shown that motion can reveal users' demographic [18, 42, 48], health [18, 47, 48], cognitive [29], and behavioral [11] information. Together, these findings show that publicly shared motion recordings may allow an attacker to infer both a user's identity and sensitive personal characteristics, placing the user's privacy at risk.

[1]BeatLeader: https://beatleader.com

## 2.2 VR Privacy Mechanisms And Their Utility Evaluations

*2.2.1 Types of VR Privacy Mechanisms.* Recognizing the privacy risks posed by VR headsets, researchers have proposed several mechanisms for protecting VR motion data. There are several types of privacy mechanisms: a) some control access to motion recordings [21], b) some propose event-based APIs that release only portions of the recordings [8], and c) some use cryptographic approaches to protect the data [20]. However, these mechanisms are impractical for public leaderboards because they a) restrict who can access the recording, effectively turning a public leaderboard into a private one, b) share only a portion of the recording, making score verification challenging, and c) may require a separate cryptographic key for each recording, limiting who can access the data and again making the leaderboard function more like a private system.

Thus, researchers have developed a specific class of privacy mechanisms for this use case: perturbation-based mechanisms [3, 16, 35, 44]. Their goal is to preserve public access to motion recordings while reducing the privacy risks associated with the data. These mechanisms operate in the following stages.

(1) **Input.** The privacy mechanism first reads the motion information contained in a VR recording. This information typically consists of time-series 3D coordinates from tracked sensors, such as the headset and controllers. For example, a recording of a user jumping would show an increase in the vertical coordinate over time.
(2) **Perturbation.** After reading the recording, the mechanism modifies the motion traces by perturbing the recorded 3D coordinates. The type of perturbation depends on the selected privacy mechanism. For example, Gaussian noise adds random jitter to the coordinates, smoothing smoothes motion trajectory, and temporal downsampling reduces the number of unique timestamps in the recording.
(3) **Output.** Finally, the perturbed motion trace is saved as a new recording. This recording contains modified 3D coordinates that differ from the original motion data. The resulting file can then be shared with others with reduced privacy risk.

The main advantage of perturbation-based mechanisms is that perturbed recordings can remain publicly accessible while the privacy risks associated with them are reduced. For example, consider a VR user who wants to upload a motion recording to BeatLeader. The user may recognize that their fast and sudden movements make their motion patterns distinctive and potentially easier to identify. To reduce this risk, they could apply a smoothing-based privacy mechanism to the recording. Smoothing may mask the abrupt movements that make the user stand out. However, perturbing the motion can also introduce adverse effects. If the recording is smoothed too aggressively, the user's movements may appear slower or delayed, potentially reducing the accuracy of the replay and affecting how their in-game performance is represented. Similar adverse effects can occur with other perturbation-based mechanisms. Gaussian noise may introduce visible jitter that reduces motion accuracy, while temporal downsampling may remove intermediate motion samples, making movements appear less smooth or causing rapid actions to be represented inaccurately. All these effect can have a negative impact on how well a mechanism is accepted by users.

*2.2.2 Utility Evaluations of VR Privacy Mechanisms.* These adverse effects introduced by privacy mechanisms (§2.2.1) motivate the need for utility evaluations of privacy mechanisms. Such evaluations help determine whether users find the mechanisms acceptable enough to use in practice. After all, a privacy mechanism provides little benefit if users are unwilling to use it. There are two common ways to conduct utility evaluations: a) distance-based evaluations [3, 16, 26, 37, 39, 40] and b) user studies [31, 43, 44]. For distance-based evaluations, prior work often reports Euclidean distance, root mean squared error, angular error, trajectory error, or related signal-fidelity measures [3, 16, 26, 37, 39, 40].

Table 1. Evaluation and participant-experience reporting practices in prior VR motion privacy mechanism work. **Distance-based** indicates utility evaluation using distance- or error-based measures such as Euclidean deviation, RMSE, trajectory error, or related signal-fidelity metrics. **User Study** indicates that participants evaluated the privacy mechanism, its usability, or the resulting protected motion. **User Study Participant Count** reports the number of participants in user-facing evaluations; papers without a user study are coded as 0. A filled circle indicates that the paper reported the corresponding measure; an open gray circle indicates that it did not.

| Paper | Year | Utility Evaluation | | User Experience Reporting | | User Study Participant Count |
|---|---|---|---|---|---|---|
| | | Distance-based | User Study | Prior VR Experience | Prior VR Privacy Mechanism Experience | |
| Ibragimov et al. [16] | 2026 | ● | ○ | ○ | ○ | 0 |
| Raju et al. [36] | 2025 | ● | ○ | ○ | ○ | 0 |
| Aziz et al. [3] | 2025 | ● | ○ | ○ | ○ | 0 |
| Aziz et al. [2] | 2025 | ○ | ○ | ○ | ○ | 0 |
| Wilson et al. [44] | 2024 | ○ | ● | ● | ○ | 18 |
| Wei et al. [43] | 2024 | ● | ● | ● | ○ | 19 |
| Ren et al. [37] | 2024 | ● | ○ | ○ | ○ | 0 |
| Nair et al. [31] | 2024 | ○ | ● | ● | ○ | 182 |
| Sun et al. [39] | 2024 | ● | ○ | ○ | ○ | 0 |
| Sun et al. [40] | 2024 | ● | ○ | ○ | ○ | 0 |
| Meng et al. [26] | 2024 | ● | ○ | ○ | ○ | 0 |
| Nair et al. [35] | 2023 | ○ | ○ | ○ | ○ | 0 |
| Hu et al. [13] | 2022 | ● | ○ | ○ | ○ | 0 |
| Moore et al. [28] | 2021 | ○ | ○ | ● | ○ | 0 |
| Gordon et al. [11] | 2021 | ○ | ○ | ○ | ○ | 0 |
| Ours | 2026 | ● | ● | ● | ● | 286 |

These metrics are useful for quantifying the amount of distortion introduced, as it is believed that the more deviation a mechanism introduces, the harder it becomes to use [16]. For user studies, researchers recruit participants to evaluate the effects of privacy mechanisms. Depending on the study design, participants may directly interact with a privacy mechanism [44], observe protected recordings or motion data [31], or evaluate its effects through other forms of presentation [43]. Researchers then collect participants' responses to understand how the mechanism and its effects are perceived by users.

However, a recent study by Wilson et al. [44] questioned the common assumption in distance-based evaluations that greater physical deviation necessarily corresponds to worse utility. They found that increasing mechanism strength (i.e., introducing greater physical deviation from the original motion) affected the usability of Gaussian noise more substantially than smoothing. In other words, the relationship between physical deviation and usability was not the same across the two mechanisms. This raises an important question for distance-based evaluations: does the amount of physical deviation introduced by a privacy mechanism correspond to how users perceive and accept the resulting motion? This relationship may also differ across users depending on their prior VR experience and exposure to VR privacy mechanisms.

To understand how prior work has addressed this question, Table 1 summarizes evaluation practices in prior work, highlighting studies that use distance-based metrics, user studies, or both. The table also reports whether these studies collected information about participants' prior experience, which may help explain differences in acceptance. Distance-based evaluations are more common, with nine papers using this approach, while user study evaluations are less common, with only three papers using this approach, and only one paper considers both. Thus, the relationship between physical deviation and users' evaluations remains unclear because prior studies rarely evaluate the same privacy mechanisms using both approaches. In this work, we directly compare physical deviation and user study evaluations to examine how well each explains privacy mechanism acceptance. We further examine whether users' prior experience affects how they evaluate different privacy mechanisms.

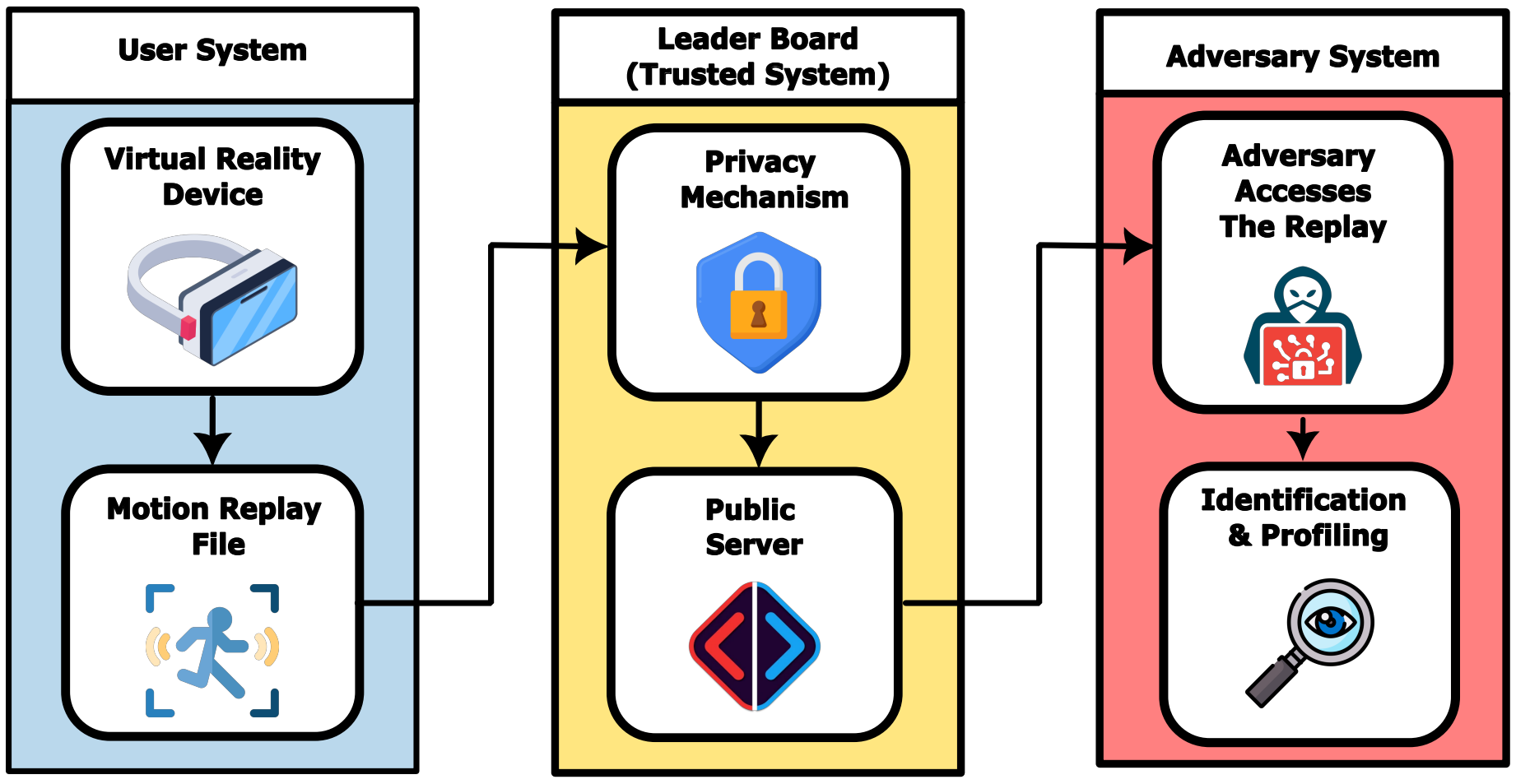

Fig. 1. **Threat model for VR motion privacy in leaderboard settings.** A user submits an original motion recording to a trusted leaderboard, which verifies the user's score and helps prevent cheating. After verification, the leaderboard applies a privacy mechanism and publishes the privatized recording. An adversary may then access the public recording and attempt to identify or profile the user.

## 3 Threat Model

We consider a VR leaderboard setting in which users submit motion recordings for score verification, after which a trusted leaderboard applies a privacy mechanism before making the recordings publicly available. Our threat model consists of three components: the system setting and motion data being protected (§3.1), the adversary and associated privacy risks (§3.2), and the role and scope of the privacy mechanism (§3.3). Figure 1 provides an overview of this setting.

### 3.1 System Setting and Protected Asset

A user plays a VR game, such as Beat Saber, and generates a motion recording containing information needed to reconstruct and verify gameplay. The recording may include head pose, hand or controller trajectories, timing information, and other movement-derived signals. After the session, the user submits the original recording to a trusted leaderboard, which uses it to verify the submitted score and help prevent cheating. Once the score is verified, the recording can be privatized and published on a public server (e.g., website, application). Prior work has shown that publicly available Beat Saber motion recordings can be used to re-identify users [30], motivating the protection of these recordings before public release.

### 3.2 Adversary Model

We assume an adversary who can access privatized motion recordings published by the leaderboard. The adversary may download or scrape these recordings, extract motion trajectories, and apply identification, linkage, or profiling models.

Their goals may include identifying or re-identifying the player, linking multiple recordings to the same user even when usernames or avatars differ, and inferring behavioral, physical, or health-related characteristics. The adversary does not need access to the user's headset, local network, or original recording; instead, the adversary operates on the privatized recording made publicly available by the leaderboard.

### 3.3 Privacy Mechanism and Scope

We consider a post-hoc privatization setting, where the leaderboard applies the privacy mechanism after verifying the user's score but before publishing the motion recording. The user first submits the original, unmodified recording, allowing the leaderboard to verify gameplay and prevent cheating. After verification, the leaderboard applies the privacy mechanism and publishes the privatized recording. Under this setting, users are not responsible for applying privacy mechanisms themselves; instead, the leaderboard is responsible for protecting the recordings it makes publicly available.

An alternative approach is real-time privatization, where a privacy mechanism modifies users' motion during gameplay. Although real-time mechanisms are commonly considered for VR motion privacy, they are not well suited to the leaderboard setting considered in this work because the leaderboard requires the original motion for score verification. If users were allowed to apply privacy mechanisms themselves during gameplay, the resulting modifications could affect gameplay-relevant motion and potentially provide an unfair advantage (e.g., extending reach, increasing movement speed, reducing motion delays, or altering controller trajectories). For this reason, we focus on post-hoc privacy mechanisms applied by the trusted leaderboard after score verification.

## 4 Method

This section describes the methodology used to investigate how users evaluate post-hoc VR motion privacy mechanisms. We first introduce the research questions (§4.1) and provide an overview of the study design (§4.2). We then describe the privacy mechanisms evaluated in the study (§4.3), the measures used to capture participants' evaluations (§4.4), and the study procedure (§4.5).

### 4.1 Research Questions

This study examines how users evaluate post-hoc VR motion privacy mechanisms in protected replay recordings. We focus on two methodological questions:

- **RQ1:** *Is measured physical deviation from the original motion trace sufficient to explain acceptance ratings of VR motion privacy mechanisms, or do user-facing perceptions explain acceptance better?*
- **RQ2:** *Do prior VR experience and prior VR privacy-mechanism experience shape acceptance ratings across mechanism type and physical deviation level?*

### 4.2 Study Design Overview

To answer the research questions, we conducted a user study in which participants reviewed and evaluated motion replay recordings after privacy mechanisms had been applied. Consistent with our threat model, we used recordings downloaded from the Beat Saber leaderboard as the base motion data for the study. These recordings contain detailed head and hand movements that are suitable for studying perceived changes in motion precision and privacy protection.

The source traces came from open-source BeatLeader replay data released by Nair et al. [30] and were visualized using the BeatLeader Analyzer Tool [5]. This tool is integrated into the BeatLeader leaderboard and allows users to analyze the movements contained in publicly available replay recordings. Example images from samples shown to participants are shown in Figure 2.

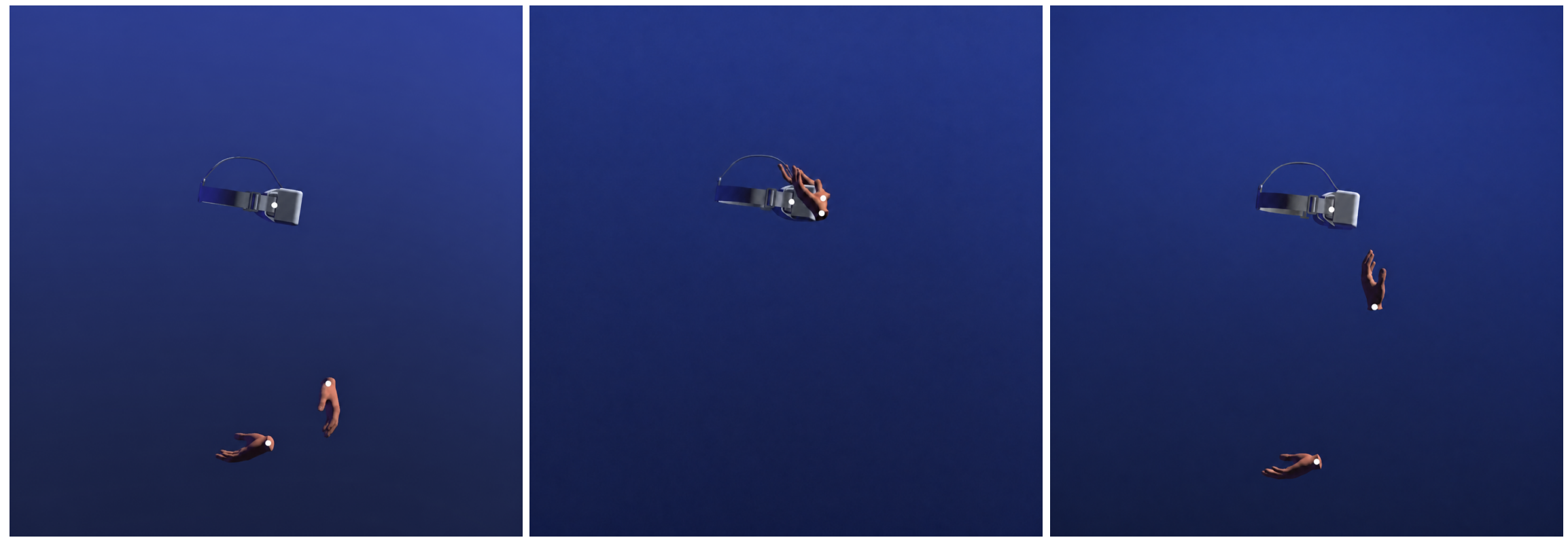

Fig. 2. Representative frames from the BeatLeader replay samples presented to participants, visualized using the BeatLeader Analyzer Tool [5]. The frames show the tracked headset and controller movements contained in the replay recordings.

The reviewed motions were influenced by two mixed-design factors: *privacy mechanism* and *level of deviation*. *Privacy mechanism* was a between-participant factor; each participant was randomly assigned to one of three mechanisms, Gaussian, Smoothing, or Temporal to all reviewed samples. *Level of deviation* was a within-participant factor; all participants evaluated five protected samples from their assigned privacy mechanism corresponding to 1, 3, 5, 7, and 11 cm of average deviation from the original motion trace. Additionally, participants self-reported their prior experience with VR and prior experience with VR privacy mechanisms. This design allowed us to compare acceptance across deviation levels, privacy mechanism types, and prior experience.

For each reviewed sample, participants answered questions grouped into five measures: perceived spatial precision, perceived temporal precision, perceived privacy risk, trust, and acceptance. These measures were motivated by prior work on VR privacy evaluation and privacy calculus models that suggest that acceptance of a privacy mechanism may depend on perceived utility, privacy protection, and trustworthiness, not only on the amount of distortion introduced by the mechanism. Further details of measures are described in Section 4.4.

### 4.3 Privacy Mechanisms

To select privacy mechanisms for the study, we focused on mechanisms that had previously been evaluated using both user studies and distance-based metrics. This provided prior evidence about both their user-facing effects and measured physical deviation. Based on this criterion, we selected three mechanisms: Gaussian [16, 33, 44], Smoothing [16, 44], and Temporal [16, 27, 44]. Prior work has evaluated these mechanisms through user studies [44] and distance-based metrics [16, 27, 33]. Other relevant mechanisms were not selected because prior work did not include either a user study [36, 37, 43] or a distance-based evaluation [13, 31, 35].

Figure 3 illustrates these effects on the same motion segment by comparing the original trajectory with each privatized signal. Let $x_t$ denote the original motion position at time $t$, and let $\tilde{x}_t$ denote the privatized position.

- **Gaussian** adds random spatial perturbations independently at each time step $\tilde{x}_t = x_t + \epsilon_t$, where $\epsilon_t$ is zero-mean Gaussian noise. As shown in Figure 3, this produces frame-to-frame jitter around the original trajectory.

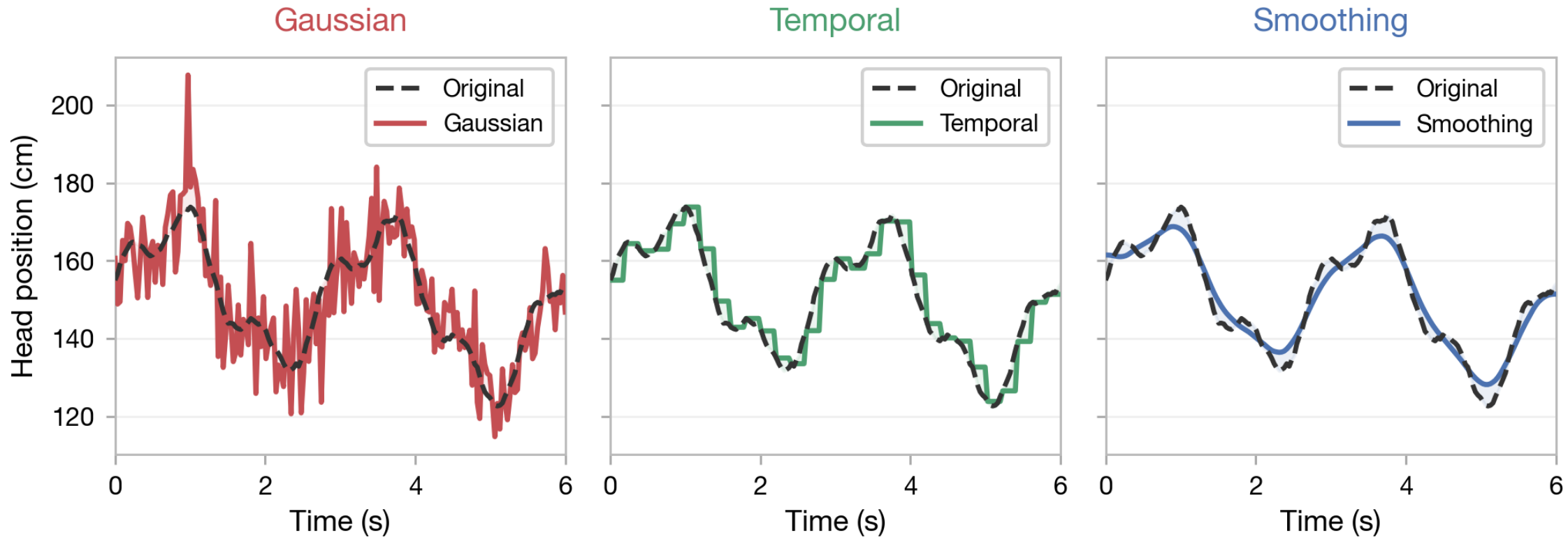

Fig. 3. Illustrative effect of three VR motion privacy mechanisms on head position over time. Each panel compares the original trajectory (dashed) with the privatized signal (solid). *Gaussian* adds random spatial noise, producing jitter. *Temporal* reduces temporal precision, producing a staircase pattern. *Smoothing* applies a moving average, attenuating rapid movements and introducing lag.

- **Smoothing** replaces each position with a local temporal average, $\tilde{x}_t = \frac{1}{K} \sum_{k=0}^{K-1} x_{t-k}$, where K is the smoothing window size and k indexes the samples within that window. In Figure 3, this attenuates rapid changes and follows the original trajectory more gradually.
- **Temporal** reduces temporal precision by holding motion values across short intervals, $\tilde{x}_t = x_{\lfloor t/K \rfloor K}$, where $K$ is the downsampling interval, measured in frames. Figure 3 shows this as a stair-step pattern, where motion freezes briefly and then jumps. This has an effect similar to reducing the system's frames per second (FPS).

Together, these mechanisms provide a useful test case for studying how users interpret privacy-preserving motion. All three reduce motion precision, but Figure 3 shows that they do so through different effects. Additionally, the same mechanism can be tuned to produce different deviation levels. Figure 4 illustrates perturbation of a head-motion signal using a Gaussian mechanism at five deviation levels used in our study. At the lowest level, the deviation is barely noticeable; however, as the deviation increases, the perturbation becomes more visible, reaching its strongest effect at 11 cm.

### 4.4 Measures

For each protected motion sample, participants rated five constructs: *perceived spatial precision*, *perceived temporal precision*, *perceived privacy risk*, *trust*, and *acceptance*. These measures were selected to capture user-facing judgments that physical deviation metrics cannot directly observe. Our motivation for these measures comes from two sources. First, prior VR motion privacy work has evaluated whether privacy mechanisms preserve the usability or perceptual quality of protected motion, suggesting that users' judgments of motion quality are important for understanding whether a mechanism is useful in practice [22, 31, 44]. Second, privacy calculus models suggest that acceptance of a privacy intervention depends on how users weigh perceived benefits, risks, and trust [9, 19]. In our setting, the relevant benefit is whether the protected replay remains usable and faithful enough to the original motion; the relevant risk is whether the motion still appears identifying; and trust reflects whether participants believe the mechanism is providing meaningful protection.

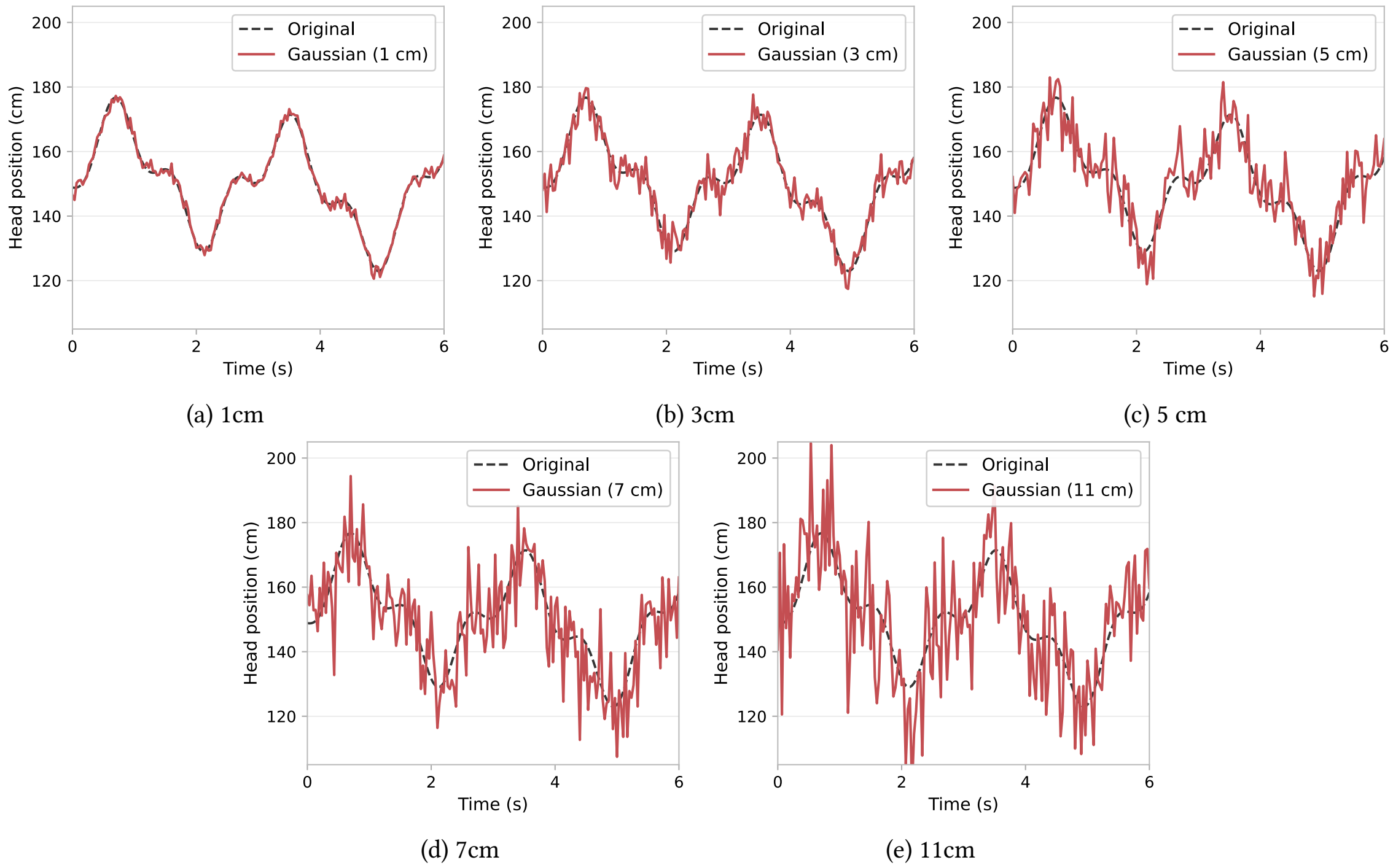

(a) 1cm (b) 3cm (c) 5 cm

(d) 7cm (e) 11cm

Fig. 4. Gaussian perturbation at five average deviation levels: (a) 1 cm, (b) 3 cm, (c) 5 cm, (d) 7 cm, and (e) 11 cm.

Figure 5 shows the item wording used for each construct. Each construct was measured with three 7-point Likert items ranging from *Strongly disagree* to *Strongly agree*, which we averaged into a construct-level score. The five constructs are described below:

- ***Perceived spatial precision*** captured whether the protected motion appeared spatially accurate. These items asked whether movements appeared precise during virtual-object interaction, whether hand movements appeared accurately represented, and whether fine motor actions appeared accurately captured. This measure reflects whether the privacy mechanism preserved the spatial quality of the replay from the participant's perspective.
- ***Perceived temporal precision*** captured whether the protected motion appeared responsive and correctly timed. These items asked whether responses appeared to occur without noticeable delay, whether movements appeared to be reflected in real time, and whether interactions appeared to occur at the intended moment. This measure reflects whether the mechanism preserved the timing and responsiveness of the replay.
- ***Perceived privacy risk*** captured residual motion-based privacy concern. These items asked whether movement patterns could still be used to identify the user, whether someone could track the user even if the username or avatar changed, and whether users with unique physical characteristics could be more easily identified from their movement patterns. This measure reflects whether participants believed the protected motion still exposed identifying information.
- ***Trust*** captured confidence that the mechanism was providing meaningful protection. These items asked whether the system appeared to actively protect movement data, whether it appeared to provide security for user movement data, and whether participants would feel confident relying on the system. This measure reflects whether participants interpreted the mechanism as a credible privacy intervention.

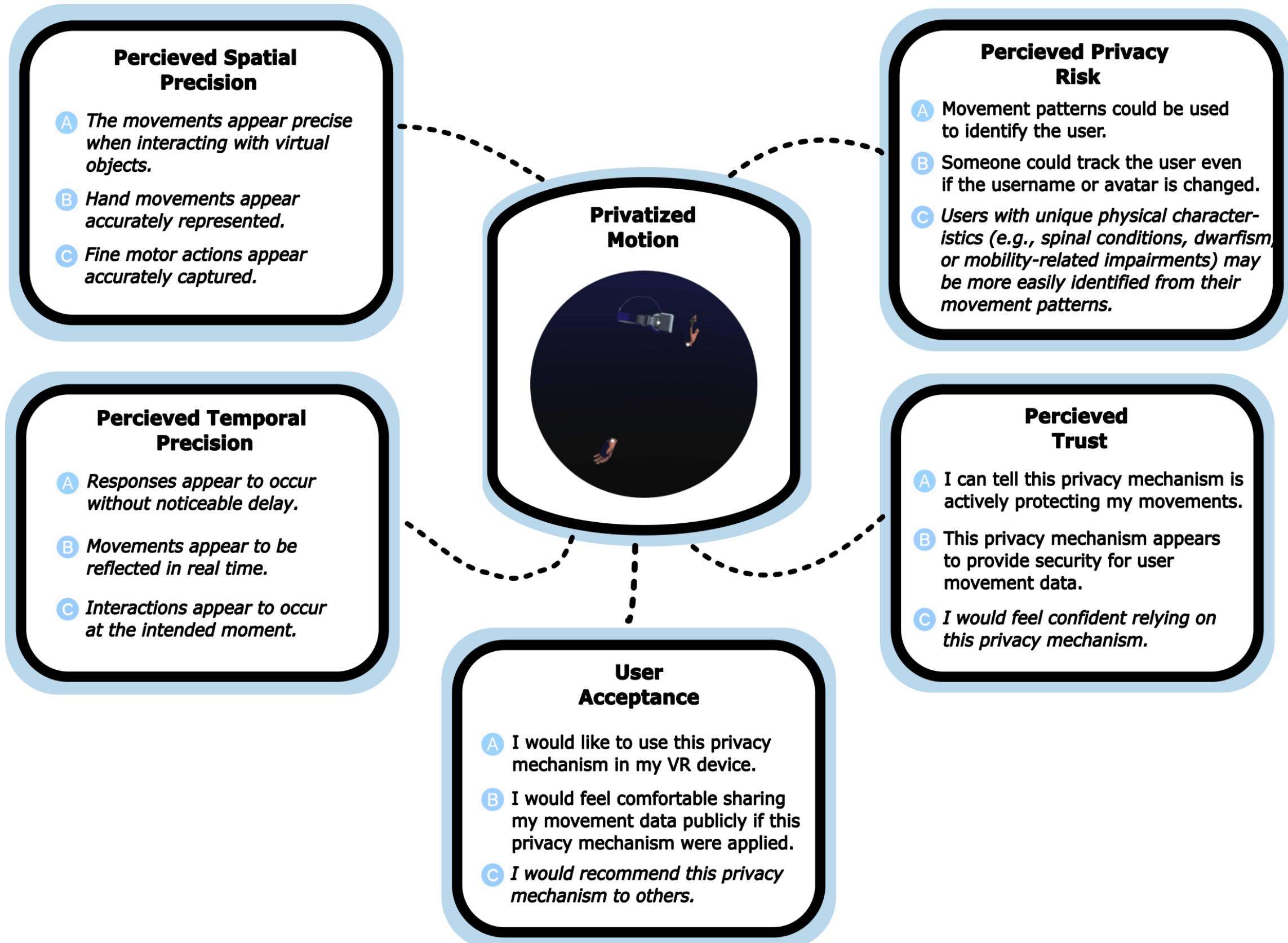

Fig. 5. Privacy mechanism rating task. For each protected motion sample, participants rated perceived spatial precision, perceived temporal precision, perceived privacy risk, trust, and acceptance. Each construct used three 7-point Likert items.

- ***Acceptance*** captured willingness to use or endorse the motion privacy mechanism. These items asked whether participants would like to use the system, would consider using it, and would recommend it to others. Acceptance served as the primary outcome because our goal was to understand whether protected motion replays were acceptable to users, not only whether the motion trace was technically transformed.

### 4.5 Procedure

This study was reviewed and approved by our university Institutional Review Board (IRB; protocol number omitted for anonymous review). Participation was limited to adults (minimum of 18 years of age), and participants provided informed consent before beginning the study. Figure 6 summarizes the study procedure. The study was conducted online via web form without synchronous researcher supervision. Participants completed the study in five stages:

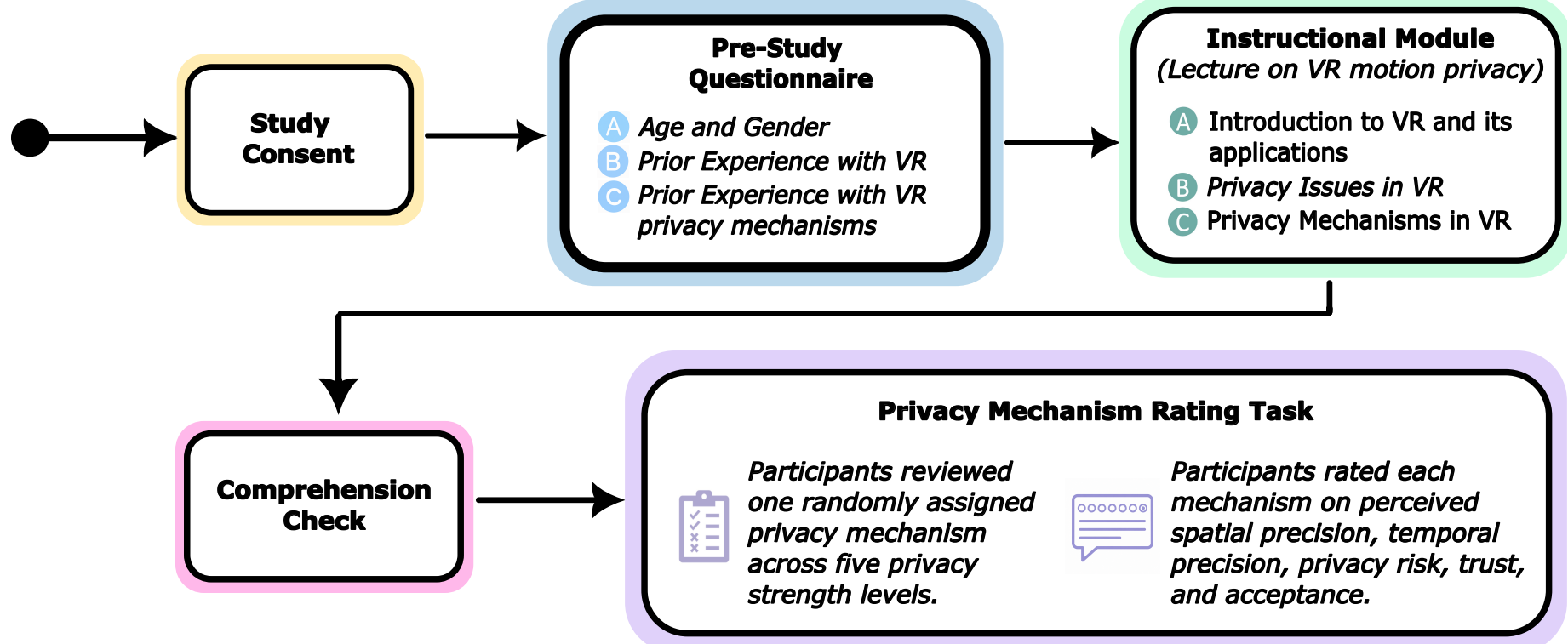

Fig. 6. Overview of the study procedure. Participants provided consent, completed a pre-study questionnaire, completed an instructional module and comprehension check, and then rated one randomly assigned privacy mechanism across five physical deviation levels.

- **Consent.** Participants first reviewed the consent form and provided informed consent before beginning the study. The study was reviewed and approved by our university Institutional Review Board (IRB; protocol number omitted for anonymous review). Eligibility was limited to participants who were at least 18 years old.
- **Pre-study questionnaire.** Participants then completed a pre-study questionnaire that collected demographic information, gaming experience, prior VR experience, and prior experience with VR privacy mechanisms. These measures were collected before participants viewed the privatized motion samples so that we could examine whether prior exposure shaped later evaluations independently of participants' responses to the rating task.
- **Instructional module.** Participants completed an instructional module on VR motion privacy. The module introduced common VR applications, explained why motion tracking is necessary for VR interaction, and described how motion traces can persist beyond gameplay through replay files, leaderboard submissions, shared recordings, and platform logs. Participants were told that the motion samples came from another person's VR gameplay recording and were instructed to evaluate them as protected replay data in a post-hoc sharing setting, where the recording could later be uploaded to a leaderboard or replay-sharing platform and viewed, ranked, verified, or analyzed by others. The module also explained that VR motion data can reveal privacy-sensitive movement patterns and introduced motion privacy mechanisms as transformations that modify recorded motion before sharing. Examples of unprivatized and privatized motion were shown so that participants could evaluate the samples as privacy interventions rather than arbitrary visual distortion.
- **Comprehension check.** After the instructional module, participants completed a comprehension check covering the core ideas from the module, including why VR motion data can be privacy-sensitive, how privacy mechanisms can modify recorded motion, and why protected replay data may still matter in sharing or leaderboard contexts. Participants could proceed to the rating task only after passing the comprehension check within three attempts. Failure to pass after three attempts would end the study for that participant and their responses were discarded.
- **Privacy mechanism rating task.** Participants were randomly assigned to one privacy mechanism, Gaussian, Smoothing, or Temporal. They then evaluated protected samples from that mechanism across the five physical deviation levels. For each sample, participants rated *perceived spatial precision*, *perceived temporal precision*, *perceived privacy risk*, *trust*, and *acceptance*. This repeated evaluation of deviation levels within a mechanism allowed us to test whether *acceptance* changed with measured physical deviation and whether user-facing perceptions explained *acceptance* beyond that deviation level. Mechanism names were not shown during the main task so that ratings reflected interpretations of protected motion rather than reactions to mechanism labels. Before the main rating task, participants completed a practice condition to become familiar with the evaluation format; practice responses were excluded from analysis.

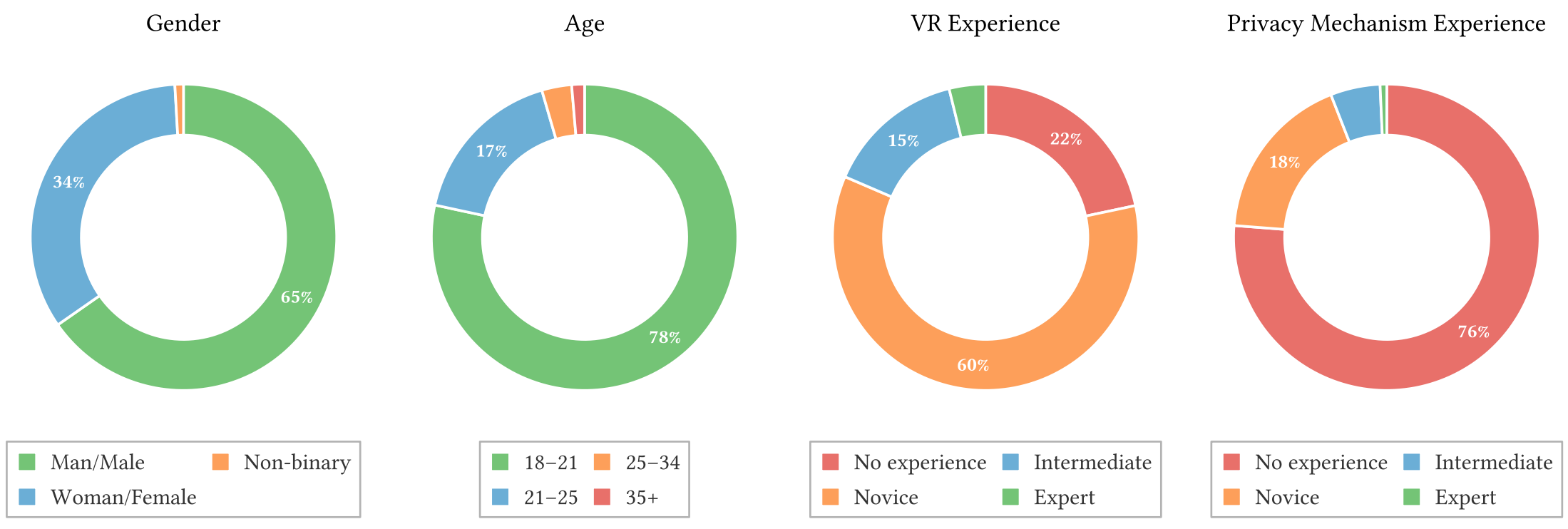

Fig. 7. Demographic characteristics and prior experience of the analysis sample ($n = 286$), including gender, age, gaming experience, VR experience, and prior VR privacy-mechanism experience.

## 5 Results and Analysis

### 5.1 Participants

We collected responses from 364 participants. After applying quality-control criteria (§5.2), the final analysis sample included 286 participants. Figure 7 summarizes demographics and prior experience for the retained sample for analysis. Participant ages ranged from 18 to 47 years, with a median age of 20. Participants were recruited from computer science courses at a university. Consistent with this course-based recruitment context, participants were mostly male (65%), primarily between 18 and 21 years of age, and generally reported some prior VR experience (78%). Prior experience with VR privacy mechanisms was less common (24%), which was expected because these mechanisms are not yet widely deployed.

### 5.2 Quality-control criteria

Participant responses were excluded if they were incomplete, showed “straightlining” [25] or had completion times below 8 minutes or above 60 minutes. The timing thresholds were selected based on the expected duration of the reasonable expected attention to study materials with time estimations conducted by the authors. To assess the minimum plausible completion time, the authors repeatedly completed the study flow while watching the instructional materials and answering the survey items as quickly as possible. Conversely, an upper bound of 60 minutes was chosen to give flexibility for slower-than-normal speeds or minor interruptions (as are common for online studies [7, 12]) while excluding larger interruptions.

### 5.3 Analysis #1: Explaining User Acceptance

To address RQ1, we developed a method that uses regression models to compare how physical deviation and user-facing ratings, separately and together, explain acceptance (§5.3.1). Using this method, we find that user-facing ratings explain substantially more variation in acceptance than physical deviation alone, while physical deviation provides a small but significant additional contribution (§5.3.2).

*5.3.1 Method of Analysis #1.* RQ1 asks whether measured physical deviation from the original motion trace is sufficient to explain acceptance ratings of VR motion privacy mechanisms, or whether user-facing perceptions explain acceptance better (§4.1). We break this question into three parts: (1) how much variation in acceptance is explained by physical deviation alone, (2) how much variation is explained by user-facing perceptions alone, and (3) whether the relative importance of these predictors remains consistent when they are considered together. To examine these relationships, we use respondent-demeaned linear regression models predicting acceptance. Respondent demeaning centers each participant's observations around their own average, allowing the models to focus on changes within participants rather than overall differences between participants. Regression allows us to estimate how each predictor (i.e., physical deviation, perceived spatial degradation, perceived temporal degradation, perceived privacy risk, and trust) is associated with acceptance and compare how much variation in acceptance is explained by different sets of predictors. Because each part requires a different predictor set, we fit three regression models:

- **Model A: Physical deviation only.** Included only measured physical deviation from the original motion trace to predict acceptance.
- **Model B: User-facing ratings only.** Included *perceived spatial degradation*, *perceived temporal degradation*, *perceived privacy risk*, and *trust* to predict acceptance.
- **Model C: Combined model.** Included both measured physical deviation and all user-facing ratings to predict acceptance.

We evaluated the models based on three aspects: (a) *explanatory power*, using within-participant explained variance ($R^2_{\text{within}}$) to determine how much variation in acceptance within participants was explained by each predictor set; (b) *predictor associations*, using standardized regression coefficients ($\beta$) to determine the direction and relative strength of each predictor's association with acceptance; and (c) *statistical significance*, using $t$ statistics and $p$ values to determine whether each association was statistically significant. A higher $R^2_{\text{within}}$ indicates that a model explains more of the changes in acceptance observed within participants, while larger absolute $\beta$ values indicate stronger associations with acceptance.

*5.3.2 Results of Analysis #1.* Figure 8 summarizes the regression models predicting acceptance. For (a) *explanatory power*, Model A, which used only measured physical deviation, explained little within-participant variation in acceptance ($R^2_{\text{within}} = .008$). Model B, which used only user-facing ratings, explained substantially more variation ($R^2_{\text{within}} = .375$). Model C, which combined physical deviation with user-facing ratings, explained the most variation ($R^2_{\text{within}} = .381$). Thus, user-facing ratings explained substantially more variation in acceptance than physical deviation alone, while adding physical deviation to the user-facing ratings provided only a small increase in explanatory power.

For (b) *predictor associations*, the coefficient patterns show that user-facing ratings were more strongly associated with acceptance than physical deviation. In Model B, trust was positively associated with *acceptance*, while *perceived spatial degradation*, *perceived temporal degradation*, and *perceived privacy risk* were negatively associated with acceptance. This pattern remained consistent in Model C after adding physical deviation. In the combined model, trust had the strongest

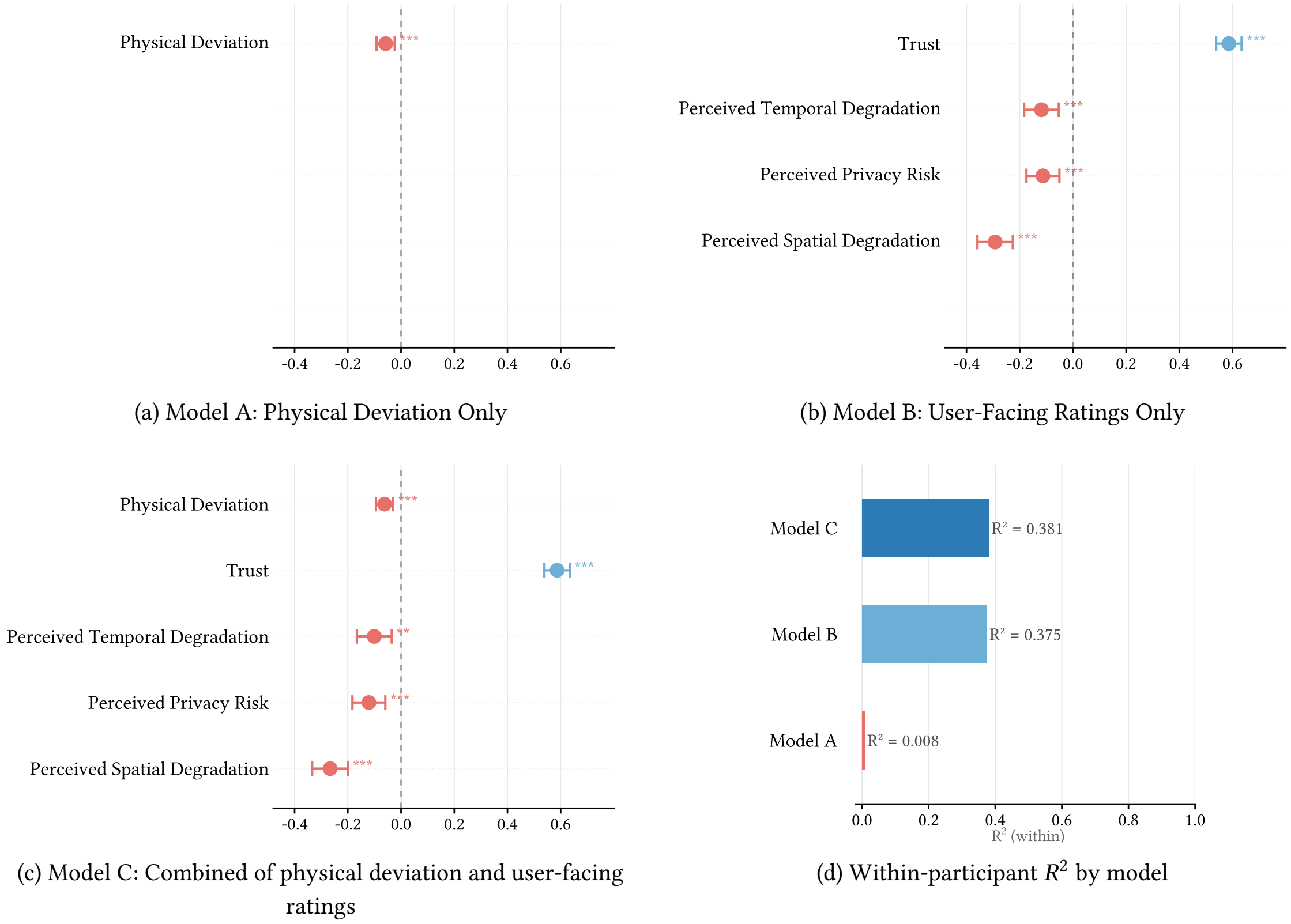

(a) Model A: Physical Deviation Only

(b) Model B: User-Facing Ratings Only

(c) Model C: Combined of physical deviation and user-facing ratings

(d) Within-participant $R^2$ by model

Fig. 8. **Analysis #1 Results.** Respondent-demeaned regression models predicting acceptance. Panels show standardized coefficients with 95% confidence intervals for (a) Model A: physical deviation only, (b) Model B: user-facing ratings only, and (c) Model C: the combined model of physical deviation and user-facing ratings. Panel (d) shows within-participant $R^2$ for each model. In panels (a)–(c), blue points indicate positive coefficients and red points indicate negative coefficients. Coefficients farther from zero indicate stronger associations with acceptance. Asterisks indicate statistical significance: $p < .05$, $p < .01$, and $p < .001$. **Overall, user-facing ratings explain substantially more variation in acceptance than physical deviation alone. Physical deviation provides limited additional explanatory power but remains significantly associated with acceptance after accounting for user-facing ratings, indicating that the two provide complementary information.**

positive association with acceptance ($\beta = 0.587$, 95% CI $[0.539, 0.634]$), followed by negative associations for *perceived spatial degradation* ($\beta = -0.267$, 95% CI $[-0.334, -0.199]$), *perceived privacy risk* ($\beta = -0.121$, 95% CI $[-0.183, -0.059]$), and *perceived temporal degradation* ($\beta = -0.100$, 95% CI $[-0.166, -0.035]$). Physical deviation had a comparatively small negative association with acceptance ($\beta = -0.062$, 95% CI $[-0.094, -0.030]$).

For (c) *statistical significance*, all predictors in the combined model were significantly associated with acceptance. Trust ($t = 24.131$, $p < .001$), perceived spatial degradation ($t = -7.733$, $p < .001$), perceived privacy risk ($t = -3.832$, $p < .001$), and perceived temporal degradation ($t = -3.013$, $p = .003$) were significant predictors. Physical deviation also remained a significant predictor ($t = -3.772$, $p < .001$), despite its relatively small standardized coefficient.

Together, these results answer RQ1 by showing that user-facing perceptions explain acceptance substantially better than measured physical deviation alone. However, these findings do not suggest that physical deviation should be replaced by user-facing evaluation. Physical deviation remained significantly associated with acceptance even after accounting for user-facing ratings, despite its relatively small standardized coefficient and the low explanatory power of physical deviation alone. Rather, the two provide complementary information for understanding acceptance. Based on these findings, we provide recommendations for evaluating VR motion privacy mechanisms in §6.1.

### 5.4 Analysis #2: Participant Experience and Sharing Acceptance

This analysis addresses RQ2, which considers whether acceptance ratings across mechanism type and physical deviation level differ based on prior experience with VR and VR privacy techniques (§4.1). We hypothesized that users may interpret the same protected motion trace differently depending on their prior experience. We developed a method to test these relationships across mechanism types and physical deviation levels (§5.4.1). Our results show that both forms of prior experience significantly influence acceptance, with their effects varying across these conditions (§5.4.2).

*5.4.1 Method of Analysis #2.* To address RQ2, we break the question into three parts: (1) whether prior VR experience and prior VR privacy-mechanism experience are associated with acceptance, (2) whether acceptance differs across mechanism types and physical deviation levels, and (3) whether experience-related differences in acceptance change across mechanism types or physical deviation levels. To examine these relationships, we use a linear mixed-effects model predicting acceptance. A mixed-effects model allows us to estimate these effects simultaneously while accounting for the repeated acceptance ratings provided by each participant.

We fit one linear mixed-effects model predicting acceptance. The model included a random intercept for participant to account for the non-independence of repeated ratings from the same participant and baseline differences in how accepting participants were overall. The fixed effects included prior VR experience, prior VR privacy-mechanism experience, mechanism type, physical deviation level, and all two-way interactions. Both experience variables were collapsed into three levels: *no experience*, *novice experience*, and *intermediate/expert experience*. We combined the highest experience categories because relatively few participants reported advanced experience, particularly with VR privacy mechanisms.

We evaluated the model based on two aspects: (a) *overall effects*, using omnibus Wald $\chi^2$ tests to determine whether each main effect or interaction was statistically significant; and (b) *group differences*, using Holm-adjusted post-hoc contrasts for statistically significant effects to identify which specific groups differed while reducing inflation from multiple pairwise comparisons. For clarity, we report only statistically significant pairwise contrasts in the results and corresponding figures.

*5.4.2 Results of Analysis #2.* Figure 9 summarizes the results of the linear mixed-effects model. For (a) *overall effects*, significant main effects were found for both *VR experience* ($\chi^2(2) = 15.43$, $p < .001$) and *VR privacy experience* ($\chi^2(2) = 10.38$, $p = .006$). In contrast, the main effects of *mechanism type* ($\chi^2(2) = 1.53$, $p = .466$) and *physical deviation level* ($\chi^2(4) = 3.30$, $p = .509$) were not statistically significant. Four two-way interactions were statistically significant: *VR experience* × *physical deviation level* ($\chi^2(8) = 31.07$, $p < .001$), *VR experience* × *mechanism type* ($\chi^2(4) = 14.28$, $p = .006$), *VR privacy experience* × *physical deviation level* ($\chi^2(8) = 16.01$, $p = .042$), and *mechanism type* × *physical deviation level* ($\chi^2(8) = 16.21$, $p = .040$). The *VR privacy experience* × *mechanism type* interaction was not statistically significant ($\chi^2(4) = 9.05$, $p = .060$).

For (b) *group differences*, the Holm-adjusted post-hoc contrasts further characterized the significant omnibus effects by identifying the specific participant groups that differed. The significant pairwise contrasts are organized below into overall experience effects, experience differences across mechanism types, and experience differences across physical deviation levels.

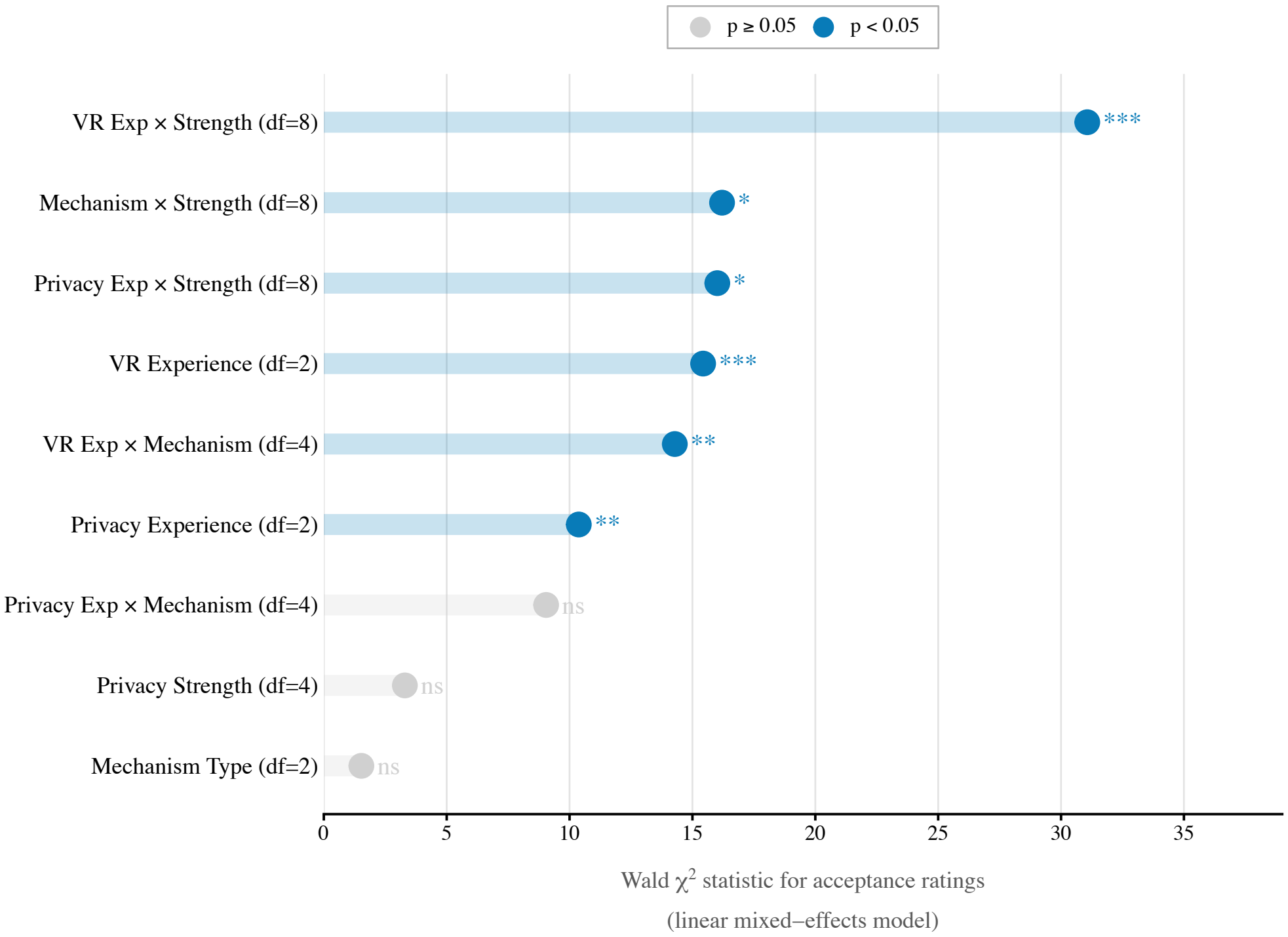

Fig. 9. **Analysis #2 Results.** Omnibus Wald $\chi^2$ tests for the **main effects** and **two-way interactions** in the linear mixed-effects model predicting sharing acceptance. Statistically significant effects were found for VR experience, VR privacy experience, and four two-way interactions, while mechanism type, physical deviation level, and the VR privacy experience × mechanism type interaction were not statistically significant. **Overall, prior VR and VR privacy-mechanism experience are significantly associated with acceptance, with these associations varying across privacy mechanisms and physical deviation levels.**

Figure 10 summarizes the significant pairwise contrasts of **acceptance ratings by experience types**. Participants with novice VR experience reported higher acceptance than participants with intermediate/expert VR experience ($p_{\text{adj}} = .039$). Participants with intermediate/expert VR privacy experience reported higher acceptance than participants with no VR privacy experience ($p_{\text{adj}} = .002$) and novice VR privacy experience ($p_{\text{adj}} = .012$).

Figure 11 summarizes the significant contrasts of **acceptance ratings across mechanism types and experience.** For the Gaussian mechanism, participants with novice VR experience reported significantly higher acceptance than participants with no VR experience ($p_{\text{adj}} = .026$) and intermediate/expert VR experience ($p_{\text{adj}} = .031$). For the Temporal mechanism, participants with no VR experience ($p_{\text{adj}} = .008$) and novice VR experience ($p_{\text{adj}} = .021$) reported significantly higher acceptance than participants with intermediate/expert VR experience. No significant differences were observed for the Smoothing mechanism.

Figure 12 summarizes the significant contrasts of **acceptance ratings across physical deviation levels and experience**. For general VR experience, participants with novice VR experience reported significantly higher acceptance than participants with no VR experience at the Very Low deviation level ($p_{\text{adj}} = .037$). At the High level, participants with novice VR experience ($p_{\text{adj}} < .001$) and no VR experience ($p_{\text{adj}} = .046$) reported significantly higher acceptance than participants with intermediate/expert VR experience. At the Very High level, participants with novice VR experience also reported significantly higher acceptance than participants with intermediate/expert VR experience ($p_{adj}$ = .010).

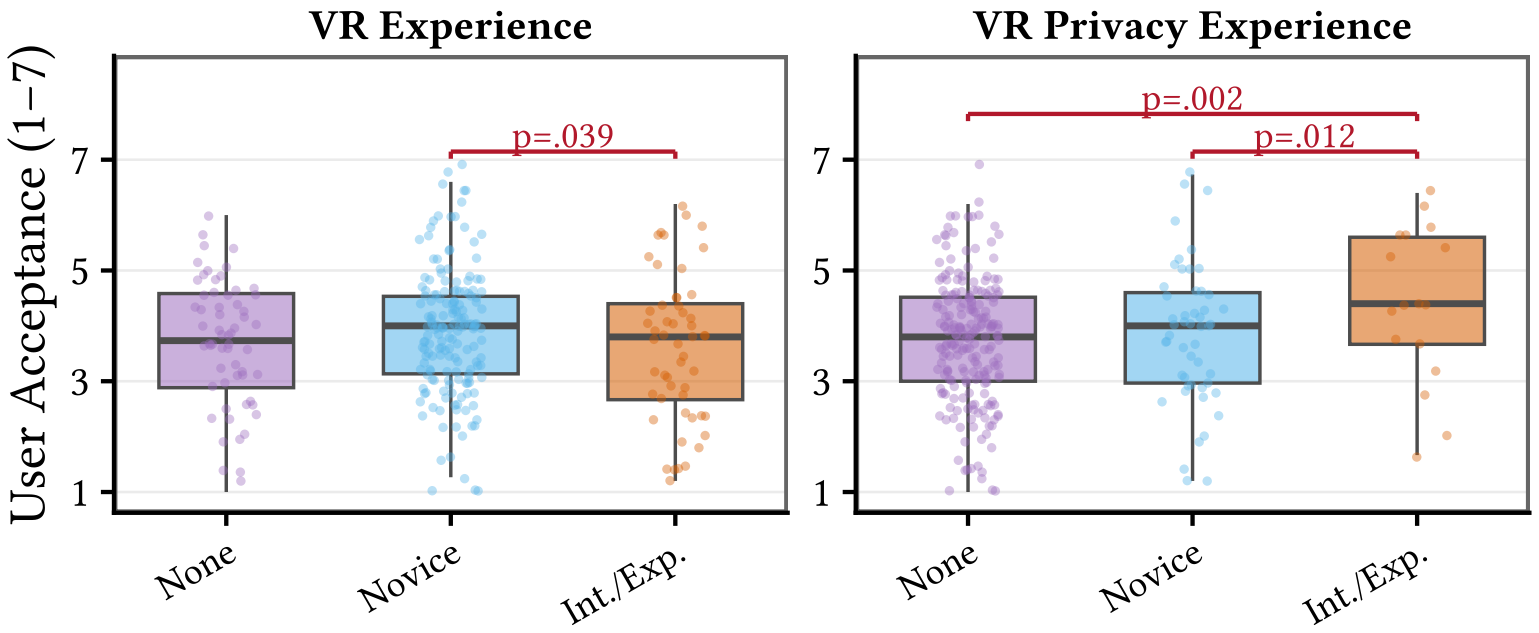

Fig. 10. Holm-adjusted pairwise contrasts for the significant **main effects** of participant experience. The figure shows estimated marginal means with 95% confidence intervals. Novice VR users reported significantly higher acceptance than intermediate/expert VR users. Participants with intermediate/expert VR privacy experience reported significantly higher acceptance than participants with no or novice VR privacy experience.

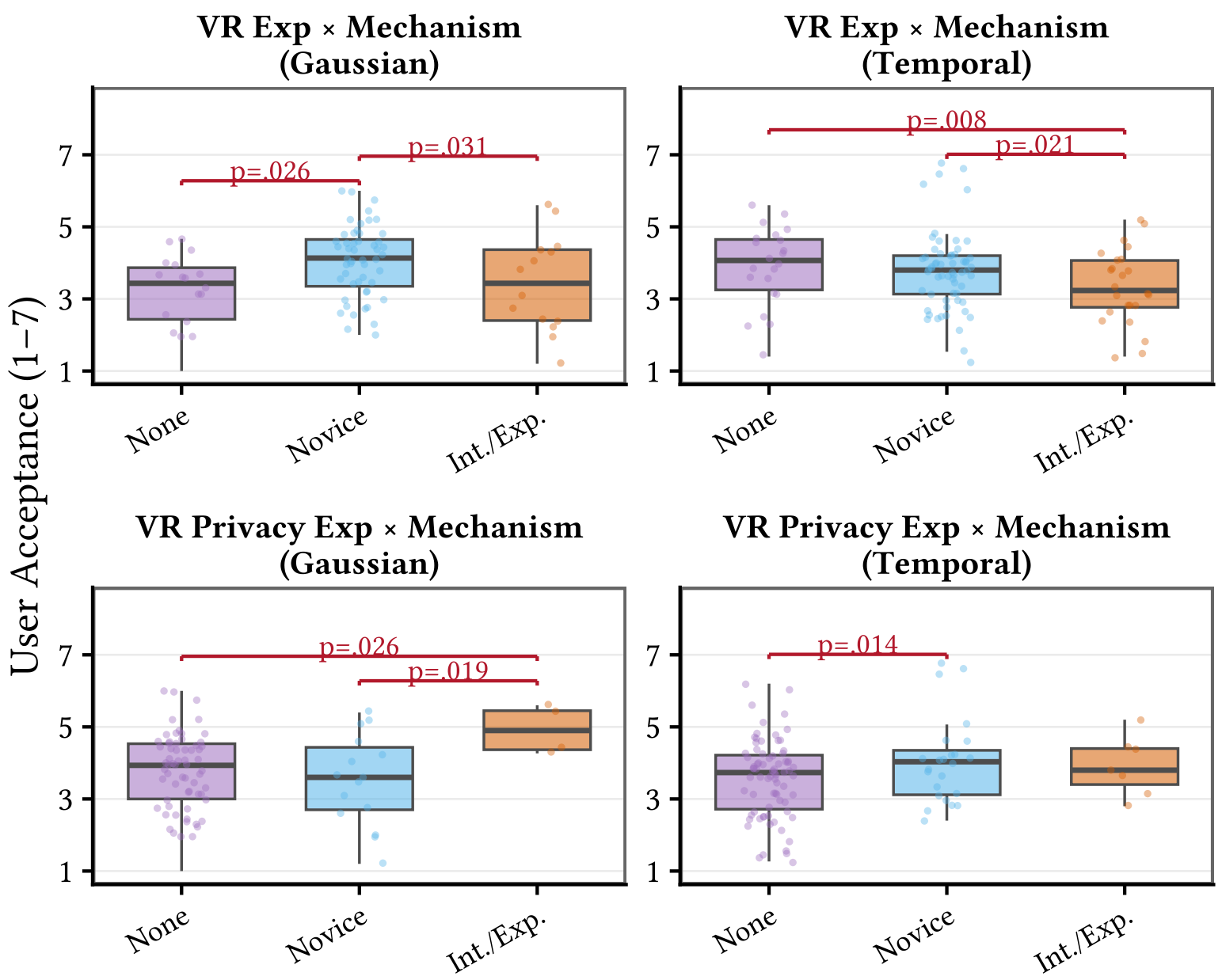

Fig. 11. Holm-adjusted pairwise contrasts **across privacy mechanisms**. The figure shows estimated marginal means with 95% confidence intervals. Significant VR-experience differences were observed for the Gaussian and Temporal mechanisms, whereas no significant differences were observed for the Smoothing mechanism. The interaction between VR privacy experience and mechanism type was not statistically significant and is therefore omitted.

For VR privacy experience, participants with intermediate/expert VR privacy experience reported significantly higher acceptance than participants with novice and no VR privacy experience at the Normal deviation level (both $p_{adj}$ < .001).

At the High level, participants with intermediate/expert VR privacy experience reported significantly higher acceptance than participants with novice VR privacy experience ($p_{\text{adj}} = .003$) and no VR privacy experience ($p_{\text{adj}} < .001$). At the Very High level, participants with intermediate/expert VR privacy experience reported significantly higher acceptance than participants with no VR privacy experience ($p_{\text{adj}} = .016$).

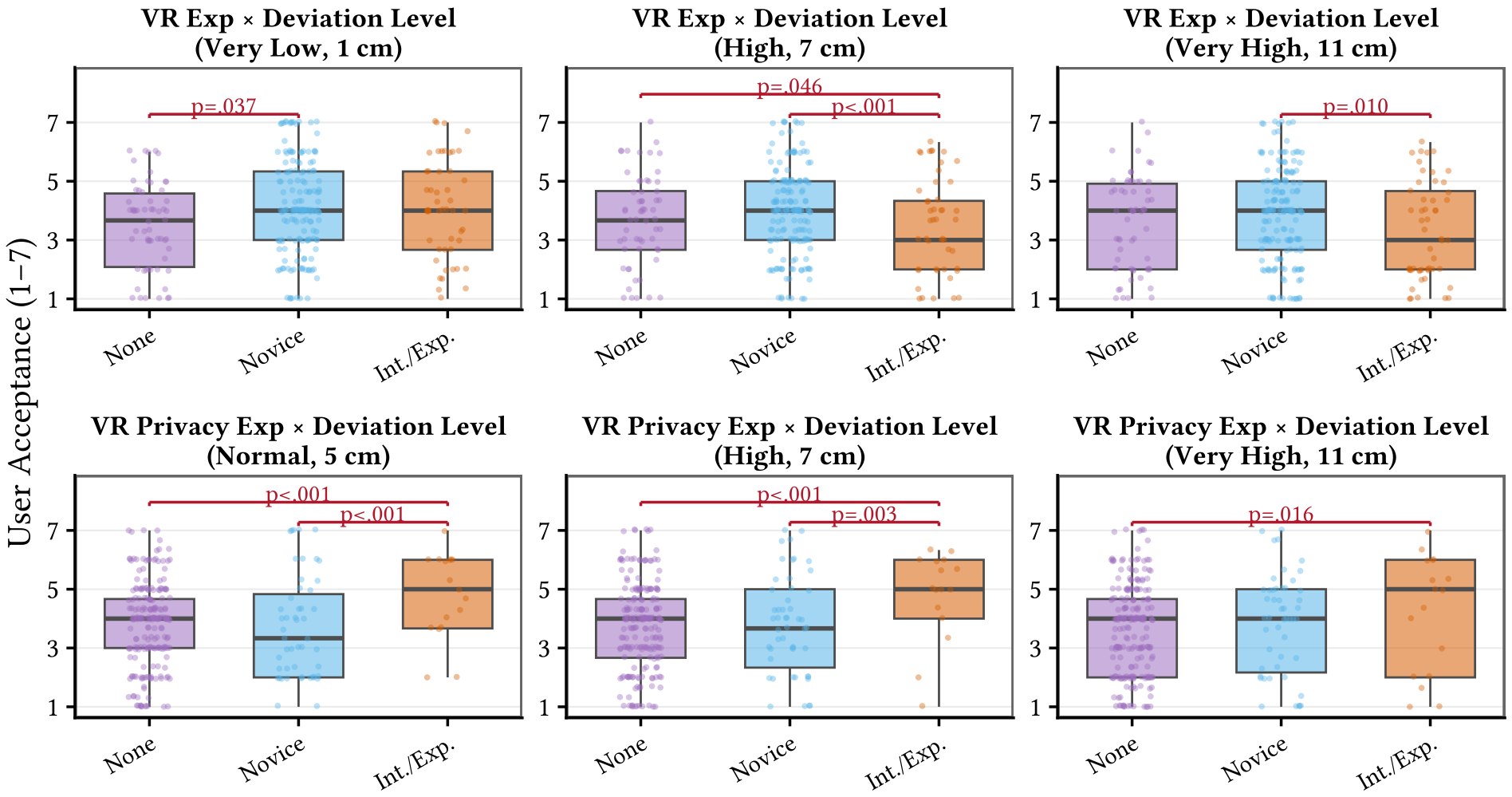

Fig. 12. Holm-adjusted pairwise contrasts across physical deviation levels. Significant differences in acceptance varied across deviation levels for both VR experience and VR privacy experience.

## 6 Discussion

The results of our study provide evidence that VR motion privacy evaluation benefits from combining distance-based, user-facing, and participant-background considerations. We first summarize the main methodological takeaways from our findings (§6.1), then discuss implications for the design and deployment of VR motion privacy mechanisms (§6.2). We next describe the limitations and contexts in which our findings should be interpreted and identify opportunities for extending these findings in future work (§6.3).

### 6.1 Key Takeaways

*Takeaway #1: Combine physical deviation with user-facing evaluations.* Physical deviation is commonly used to evaluate the utility of VR motion privacy mechanisms because it provides an objective measure of how much protected motion differs from the original motion. However, the same amount of physical deviation can produce different perceptual effects depending on the privacy mechanism. Our results show that physical deviation was significantly associated with sharing acceptance, but user-facing ratings explained substantially more variation in acceptance than physical deviation alone. Importantly, physical deviation remained significantly associated with acceptance even after accounting for user-facing ratings, indicating that it provides complementary information beyond these user-facing evaluations. Thus, our findings do not suggest replacing physical deviation with user-facing evaluations. Rather, they show that physical deviation provides useful objective information but, on its own, captures only a limited part of how users evaluate protected motion for sharing. We therefore recommend evaluating VR motion privacy mechanisms using both physical deviation and user-facing measures rather than relying on either alone.

*Takeaway #2: Consider participant experience when evaluating VR privacy mechanisms.* Results from Analysis #2 showed that participants' prior experience with VR and VR privacy mechanisms was associated with sharing acceptance across mechanism types and physical deviation levels. General VR experience and VR privacy-mechanism experience showed different patterns. Novice VR users reported higher acceptance than intermediate/expert VR users in several comparisons, suggesting that more experienced VR users may be more sensitive to changes in protected motion. In contrast, participants with greater VR privacy-mechanism experience reported higher acceptance than less experienced participants, particularly at moderate-to-high deviation levels. This may suggest greater tolerance for visible changes when participants have prior experience with privacy mechanisms and understand their protective purpose. Together, these findings show that participants' prior experience may affect how they evaluate protected motion. We therefore recommend that user studies of VR privacy mechanisms collect and report participants' prior VR and VR privacy-mechanism experience and consider this experience when interpreting study results.

### 6.2 Design Implications for VR Replay-Sharing and Leaderboard Platforms

Our study was motivated by VR replay-sharing and leaderboard platforms such as BeatLeader, where users upload motion-rich gameplay recordings that can later be accessed by anyone. As described in our threat model (§3), a leaderboard could verify the original replay, apply a privacy mechanism, and then publish only the protected replay. Our findings provide several implications for how these platforms could incorporate privacy protection into the replay-sharing process.

- **Allow users to preview protected replays before they are published.** The study results show that physical deviation alone may not capture whether users find protected motion acceptable. Two protected replays with similar physical deviations can change motion in different ways, and users may perceive these changes differently. This is particularly important for leaderboard platforms because the protected replay becomes the public representation of how the user performed. Rather than automatically applying a privacy mechanism and immediately publishing the resulting replay, a platform such as BeatLeader could allow users to preview the protected replay before it becomes public. For example, after verifying the original replay, the platform could generate several protected versions using different privacy mechanisms or protection levels and allow the user to view how each version represents their movements. The user could then select the version they are comfortable sharing publicly. This would allow the leaderboard to maintain privacy protection while giving users an opportunity to evaluate how that protection changes the public representation of their performance.
- **Configure privacy protection using properties that users can directly evaluate.** A leaderboard must also decide how users select the amount or type of privacy protection applied to their replay. One option would be to expose physical deviation directly, for example, by asking users to choose between 1cm, 3cm, or 5 cm of deviation. However, our study showed that physical deviation explained substantially less variation in sharing acceptance than users' perceptions of the protected motion. This suggests that physical deviation may be useful for configuring the privacy mechanism internally, but may not be the most meaningful way to present privacy choices to users. Instead, a leaderboard could allow users to express preferences based on properties they can directly evaluate, such as preserving the appearance of their movements, preserving the timing and smoothness of the replay, or prioritizing stronger privacy protection even when it produces more noticeable changes. The platform could then use these preferences to select an appropriate privacy mechanism and strength. For example, a user who cares strongly about preserving motion timing could receive a configuration that introduces less temporal modification, while a user who prioritizes privacy protection could receive a stronger transformation. In this way, physical deviation can remain an underlying technical measure used by the platform while the privacy choices presented to users reflect properties that are more closely related to how they evaluate protected motion.

Together, these implications suggest that privacy protection could become part of the replay-sharing process itself. After verifying the original recording, a leaderboard could generate protected versions, allow the user to evaluate how their motion will appear publicly, and publish the protected replay selected by the user. This approach preserves the leaderboard's ability to use the original recording for score verification while giving users greater control over how their motion is represented once the replay becomes publicly accessible.

### 6.3 Limitations and Future Work

Our findings identify several opportunities for future work that were either a) outside our leaderboard threat model or b) difficult to examine given the current deployment of VR privacy mechanisms.

- **Design privacy mechanisms around users' perceptions.** Prior work has shown that, within a privacy mechanism, increasing physical deviation can reduce identification risk and provide stronger privacy protection [16]. Our results show that users' perceptions of protected motion explained substantially more variation in sharing acceptance than physical deviation alone. Together, these findings suggest that it may be possible to introduce greater physical deviations, and therefore provide stronger privacy protection, while changing motion in ways that users still find acceptable. Future work could explore this possibility by designing privacy mechanisms around how different motion changes are perceived and using user studies to identify transformations that provide stronger privacy while maintaining acceptable representations of users' motion.
- **Study users with greater privacy-mechanism experience.** Because VR privacy mechanisms are still an emerging research technology and are not widely deployed, populations with substantial prior experience are currently difficult to identify. We therefore recruited from a university context where several prior studies involving VR privacy mechanisms had been conducted, providing a population with greater opportunities for prior exposure than would typically be available. Even in this setting, only 24% of participants reported prior privacy-mechanism experience, and participants with higher levels of experience remained uncommon. Future work could address this limitation through longitudinal studies that repeatedly expose participants to VR privacy mechanisms, allowing researchers to examine how users' evaluations change as they gain experience and familiarity over time.
- **Evaluate different threat models.** Our study focused on the leaderboard threat model, where publicly available motion recordings can expose large populations of users to identification attacks. Prior work has demonstrated this risk at the scale of more than 55,000 users, making public leaderboards a particularly scalable and immediate setting for studying VR motion privacy. However, other threat models should also be explored, including multiplayer, social, and casual VR applications. These settings may introduce different privacy risks and requirements for privacy mechanisms. For example, real-time privacy mechanisms were not suitable for our leaderboard threat model because the original motion is needed to verify users' performance, but they may be more appropriate when protecting motion shared with other users during active VR interaction. Future studies could examine whether our findings extend to these threat models and how users evaluate privacy mechanisms when protection occurs during real-time VR use.

## 7 Conclusion

This work presented a controlled user study with 286 participants evaluating three VR motion privacy mechanisms across five physical deviation levels (§4). The first analysis compared distance-based and user-facing evaluations for understanding sharing acceptance of protected VR motion (§5.3). The results show that both physical deviation and user-facing perceptions are significantly associated with sharing acceptance, but user-facing perceptions explain substantially more variation in acceptance than physical deviation alone. The second analysis examined whether prior VR experience and prior exposure to VR privacy mechanisms shape sharing acceptance across mechanism types and physical deviation levels (§5.4). The results show that participants' prior experience significantly affects sharing acceptance. Based on these findings, the paper recommends combining distance-based metrics with user-facing evaluations and collecting and reporting participants' prior experience when evaluating VR privacy mechanisms (§6.1). These findings also motivate design implications for replay-sharing and leaderboard platforms, including giving users greater visibility into and control over how protected motion is represented before sharing (§6.2). Finally, the limitations of the study identify important contexts in which these findings should be interpreted and directions for extending them across broader populations, applications, and privacy mechanisms (§6.3).

## 8 Open Science

To support transparency and reproducibility, we release the study materials presented to participants. Specifically, we provide the instructional module used to introduce participants to VR motion privacy, as well as the video stimuli used in the privacy mechanism rating task. These materials are included as supplementary material with this submission.

## References

[1] ALT Lab VR. [n. d.]. Echo VR. https://www.altlabvr.com/echo-vr. Accessed: 2026-09-02.

[2] Samantha Aziz and Oleg Komogortsev. 2025. Exploring the Uncoordinated Privacy Protections of Eye Tracking and VR Motion Data for Unauthorized User Identification. In *2025 IEEE Conference Virtual Reality and 3D User Interfaces (VR)*. 217–227. doi:10.1109/VR59515.2025.00046

[3] Samantha Aziz and Oleg Komogortsev. 2025. Privacy Enhancement for Gaze Data Using a Noise-Infused Autoencoder. In *2025 IEEE International Joint Conference on Biometrics (IJCB)*. 1–10. doi:10.1109/IJCB65343.2025.11411516

[4] Beat Games. [n. d.]. Beat Saber. https://beatsaber.com/. Accessed: 2026-09-02.

[5] BeatLeader. 2026. BL Replay Analyzer. https://analyzer.beatleader.com. Accessed: 2026-06-01.

[6] Buckethead Entertainment. [n. d.]. RUMBLE. https://store.steampowered.com/app/890550/RUMBLE/. Accessed: 2026-09-02.

[7] Scott Clifford and Jennifer Jerit. 2014. Is there a cost to convenience? An experimental comparison of data quality in laboratory and online studies. *Journal of Experimental Political Science* 1, 2 (2014), 120–131. doi:10.1017/xps.2014.5

[8] Brendan David-John, Diane Hosfelt, Kevin Butler, and Eakta Jain. 2021. A privacy-preserving approach to streaming eye-tracking data. *IEEE Transactions on Visualization and Computer Graphics* 27, 5 (2021), 2555–2565. doi:10.1109/TVCG.2021.3067787

[9] Elizabeth Fife and Juan Orjuela. 2012. The privacy calculus: Mobile apps and user perceptions of privacy and security. *International Journal of Engineering Business Management* 4 (2012), 11. doi:10.5772/51645

[10] Google. [n. d.]. Tilt Brush. https://store.steampowered.com/app/327140/Tilt_Brush/. Accessed: 2026-09-02.

[11] Jeremy Raboff Gordon, Max T. Curran, John Chuang, and Coye Cheshire. 2021. Covert Embodied Choice: Decision-Making and the Limits of Privacy Under Biometric Surveillance. In *Proceedings of the 2021 CHI Conference on Human Factors in Computing Systems* (Yokohama, Japan) *(CHI '21)*. Association for Computing Machinery, New York, NY, USA, Article 551, 12 pages. doi:10.1145/3411764.3445309

[12] Sandy JJ Gould, Anna L Cox, Duncan P Brumby, and Sarah Wiseman. 2015. Home is where the lab is: A comparison of online and lab data from a time-sensitive study of interruption. *Human Computation* 2, 1 (2015), 45–67. doi:10.15346/hc.v2i1.4

[13] Miao Hu, Zhenxiao Luo, Yipeng Zhou, Xuezheng Liu, and Di Wu. 2022. Otus: A Gaze Model-based Privacy Control Framework for Eye Tracking Applications. In *IEEE INFOCOM 2022 - IEEE Conference on Computer Communications*. 560–569. doi:10.1109/INFOCOM48880.2022.9796665

[14] Azim Ibragimov, Mauricio Pamplona Segundo, Sudeep Sarkar, and Kevin Bowyer. 2024. Unveiling Gender Effects in Gait Recognition Using Conditional-Matched Bootstrap Analysis. In *2024 IEEE 18th International Conference on Automatic Face and Gesture Recognition (FG)*. 1–10. doi:10.1109/FG59268.2024.10582036

[15] Azim Ibragimov, Alina Vasina, Uliana Polshcha, and Eric D. Ragan. 2026. SoK: Motion Data Privacy in Extended Reality. arXiv:2609.00711 [cs.CR] https://arxiv.org/abs/2609.00711

[16] Azim Ibragimov, Ethan Wilson, Kevin R. B. Butler, and Eakta Jain. 2026. Toward Multimodal Privacy in XR: Design and Evaluation of Composite Privatization Methods for Gaze and Body Tracking Data. *IEEE Transactions on Visualization and Computer Graphics* 32, 5 (2026), 4396–4407. doi:10.1109/TVCG.2026.3679093

[17] Anil K Jain, Patrick Flynn, and Arun A Ross. 2007. *Handbook of biometrics*. Springer Science & Business Media.

[18] Yeonju Jang, Jose A. Guridi, Daniel Molitor, Rachel Kwon, and Sneha Nagarajan. 2023. Are You Anonymous? Inferring Personal Information from Nonverbal Behavior Data Tracked in Immersive Virtual Reality. In *Companion Publication of the 2023 Conference on Computer Supported Cooperative Work and Social Computing* (Minneapolis, MN, USA) *(CSCW '23 Companion)*. Association for Computing Machinery, New York, NY, USA, 172–176. doi:10.1145/3584931.3607004

[19] Mark J. Keith, Samuel C. Thompson, Joanne Hale, Paul Benjamin Lowry, and Chapman Greer. 2013. Information disclosure on mobile devices: Re-examining privacy calculus with actual user behavior. *International Journal of Human-Computer Studies* 71, 12 (2013), 1163–1173. doi:10.1016/j.ijhcs.2013.08.016

[20] Roop Kumar Yekollu, Tejal Bhimraj Ghuge, Sammip Sunil Biradar, Shivkumar V Haldikar, and Omer Farook Mohideen Abdul Kader. 2024. Securing the Virtual Realm: Strategies for Cybersecurity in Augmented Reality (AR) and Virtual Reality (VR) Applications. In *2024 8th International Conference on I-SMAC (IoT in Social, Mobile, Analytics and Cloud) (I-SMAC)*. 520–526. doi:10.1109/I-SMAC61858.2024.10714591

[21] Karina LaRubbio, Ethan Wilson, Sanjeev Koppal, Sophie Jörg, and Eakta Jain. 2023. Give me some room please! Personal space bubbles for safety and performance. In *2023 IEEE Conference on Virtual Reality and 3D User Interfaces Abstracts and Workshops (VRW)*. 897–898. doi:10.1109/VRW58643.2023.00291

[22] Jingjie Li, Amrita Roy Chowdhury, Kassem Fawaz, and Younghyun Kim. 2021. Kaleido: Real-Time privacy control for {Eye-Tracking} systems. In *30th USENIX security symposium (USENIX security 21)*. 1793–1810.

[23] Xiang Li, Yasushi Makihara, Chi Xu, Yasushi Yagi, and Mingwu Ren. 2020. Gait Recognition via Semi-supervised Disentangled Representation Learning to Identity and Covariate Features. In *2020 IEEE/CVF Conference on Computer Vision and Pattern Recognition (CVPR)*. 13306–13316. doi:10.1109/CVPR42600.2020.01332

[24] Yasushi Makihara, Atsuyuki Suzuki, Daigo Muramatsu, Xiang Li, and Yasushi Yagi. 2017. Joint Intensity and Spatial Metric Learning for Robust Gait Recognition. In *2017 IEEE Conference on Computer Vision and Pattern Recognition (CVPR)*. 6786–6796. doi:10.1109/CVPR.2017.718

[25] Adam W Meade and S Bartholomew Craig. 2012. Identifying careless responses in survey data. *Psychological methods* 17, 3 (2012), 437.

[26] Yan Meng, Yuxia Zhan, Jiachun Li, Suguo Du, Haojin Zhu, and Xuemin Shen. 2024. De-anonymizing avatars in virtual reality: Attacks and countermeasures. *IEEE Transactions on Mobile Computing* 23, 12 (2024), 13342–13357.

[27] Mark Roman Miller, Vivek Nair, Eugy Han, Cyan DeVeaux, Christian Rack, Rui Wang, Brandon Huang, Marc Erich Latoschik, James F O'Brien, and Jeremy N Bailenson. 2024. Effect of Duration and Delay on the Identifiability of VR Motion. In *2024 IEEE 25th International Symposium on a World of Wireless, Mobile and Multimedia Networks (WoWMoM)*. IEEE, 70–75.

[28] Alec G. Moore, Ryan P. McMahan, Hailiang Dong, and Nicholas Ruozzi. 2021. Personal Identifiability and Obfuscation of User Tracking Data From VR Training Sessions. In *2021 IEEE International Symposium on Mixed and Augmented Reality (ISMAR)*. 221–228. doi:10.1109/ISMAR52148.2021.00037

[29] John E Munoz, Faraz Ali, Aysha Basharat, Samira Mehrabi, Michael Barnett-Cowan, Shi Cao, Laura E Middleton, and Jennifer Boger. 2023. Development of classifiers to determine factors associated with older adult's cognitive functions and game user experience in VR using head kinematics. *IEEE Transactions on Games* 17, 3 (2023), 604–612.

[30] Vivek Nair, Wenbo Guo, Justus Mattern, Rui Wang, James F O'Brien, Louis Rosenberg, and Dawn Song. 2023. Unique identification of 50,000+ virtual reality users from head & hand motion data. In *32nd USENIX Security Symposium (USENIX Security 23)*. 895–910.

[31] Vivek Nair, Wenbo Guo, James F. O'Brien, Louis Rosenberg, and Dawn Song. 2024. Deep Motion Masking for Secure, Usable, and Scalable Real-Time Anonymization of Ecological Virtual Reality Motion Data. In *2024 IEEE Conference on Virtual Reality and 3D User Interfaces Abstracts and Workshops (VRW)*. 493–500. doi:10.1109/VRW62533.2024.00096

[32] Vivek Nair, Wenbo Guo, Rui Wang, James F O'Brien, Louis Rosenberg, and Dawn Song. 2024. Berkeley Open Extended Reality Recordings 2023 (BOXRR-23): 4.7 Million Motion Capture Recordings from 105,000 XR Users. *IEEE Transactions on Visualization and Computer Graphics* 30, 5 (2024), 2239–2246. doi:10.1109/TVCG.2024.3372087

[33] Vivek Nair, Mark Roman Miller, Rui Wang, Brandon Huang, Christian Rack, Marc Erich Latoschik, and James F O'Brien. 2024. Effect of data degradation on motion re-identification. In *2024 IEEE 25th International Symposium on a World of Wireless, Mobile and Multimedia Networks (WoWMoM)*. IEEE, 85–90.

[34] Vivek Nair, Louis Rosenberg, James F. O'Brien, and Dawn Song. 2024. Truth in Motion: The Unprecedented Risks and Opportunities of Extended Reality Motion Data. *IEEE Security & Privacy* 22, 1 (2024), 24–32. doi:10.1109/MSEC.2023.3330392

[35] Vivek C Nair, Gonzalo Munilla-Garrido, and Dawn Song. 2023. Going Incognito in the Metaverse: Achieving Theoretically Optimal Privacy-Usability Tradeoffs in VR. In *Proceedings of the 36th Annual ACM Symposium on User Interface Software and Technology* (San Francisco, CA, USA) *(UIST '23)*. Association for Computing Machinery, New York, NY, USA, Article 61, 16 pages. doi:10.1145/3586183.3606754

[36] Mehedi Hasan Raju and Oleg V Komogortsev. 2025. Real-Time Lightweight Gaze Privacy-Preservation Techniques Validated via Offline Gaze-Based Interaction Simulation. *arXiv preprint arXiv:2511.09846* (2025).

[37] Xiaojun Ren, Jiluan Fan, Ning Xu, Shaowei Wang, Changyu Dong, and Zikai Wen. 2024. DPGazeSynth: Enhancing eye-tracking virtual reality privacy with differentially private data synthesis. *Information Sciences* 675 (2024), 120720. doi:10.1016/j.ins.2024.120720

[38] Reza Shahriari and Eric D. Ragan. 2026. A Systematic Survey of Empirical User Studies of Unintentional Information Disclosure in Everyday Digital Interaction. *International Journal of Human–Computer Interaction* 0, 0 (2026), 1–29. arXiv:https://doi.org/10.1080/10447318.2026.2620054 doi:10.1080/10447318.2026.2620054

[39] Ruoxi Sun, Hanwen Wang, Hsiang-Ting Chen, and Minhui Xue. 2025. Privacy in Motion: Implementing Differential Privacy for User Motion in VR. In *Proceedings of the 36th Australasian Conference on Human-Computer Interaction (OzCHI '24)*. Association for Computing Machinery, New York, NY, USA, 223–231. doi:10.1145/3726986.3727017

[40] Ruoxi Sun, Hanwen Wang, Minhui Xue, and Hsiang-Ting Chen. 2024. PPVR: A Privacy-Preserving Approach for User Behaviors in VR. In *2024 IEEE Conference on Virtual Reality and 3D User Interfaces Abstracts and Workshops (VRW)*. 1055–1056. doi:10.1109/VRW62533.2024.00324

[41] Emmeline Taylor. 2010. I Spy with My Little Eye: The Use of CCTV in Schools and the Impact on Privacy. *The Sociological Review* 58, 3 (2010), 381–405. arXiv:https://doi.org/10.1111/j.1467-954X.2010.01930.x doi:10.1111/j.1467-954X.2010.01930.x

[42] PP Tricomi, F Nenna, L Pajola, M Conti, and L Gamberini. 2023. You can't hide behind your headset: user profiling in augmented and virtual reality. IEEE Access 11: 9859–9875. *Online: https://doi. org/10.1109/ACCESS* (2023).

[43] Yu-Szu Wei, Yuan-Chun Sun, Shin-Yi Zheng, Hsun-Fu Hsu, Chun-Ying Huang, and Cheng-Hsin Hsu. 2024. Mitigating Privacy Threats Without Degrading Visual Quality of VR Applications: Using Re-Identification Attack as a Case Study. In *2024 IEEE 7th International Conference on Multimedia Information Processing and Retrieval (MIPR)*. IEEE, 214–220.

[44] Ethan Wilson, Azim Ibragimov, Michael J. Proulx, Sai Deep Tetali, Kevin Butler, and Eakta Jain. 2024. Privacy-Preserving Gaze Data Streaming in Immersive Interactive Virtual Reality: Robustness and User Experience. *IEEE Transactions on Visualization and Computer Graphics* 30, 5 (2024), 2257–2268. doi:10.1109/TVCG.2024.3372032

[45] Chi Xu, Shogo Tsuji, Yasushi Makihara, Xiang Li, and Yasushi Yagi. 2023. Occluded Gait Recognition via Silhouette Registration Guided by Automated Occlusion Degree Estimation. In *2023 IEEE/CVF International Conference on Computer Vision Workshops (ICCVW)*. 3191–3201. doi:10.1109/ICCVW60793.2023.00344

[46] Kaihao Zhang, Wenhan Luo, Lin Ma, Wei Liu, and Hongdong Li. 2019. Learning Joint Gait Representation via Quintuplet Loss Minimization. In *2019 IEEE/CVF Conference on Computer Vision and Pattern Recognition (CVPR)*. 4695–4704. doi:10.1109/CVPR.2019.00483

[47] Tianfang Zhang, Cong Shi, Tianming Zhao, Zhengkun Ye, Payton Walker, Nitesh Saxena, Yan Wang, and Yingying Chen. 2022. Personalized health monitoring via vital sign measurements leveraging motion sensors on AR/VR headsets. In *Proceedings of the 20th Annual International Conference on Mobile Systems, Applications and Services* (Portland, Oregon) *(MobiSys '22)*. Association for Computing Machinery, New York, NY, USA, 529–530. doi:10.1145/3498361.3538768

[48] Tianfang Zhang, Zhengkun Ye, Ahmed Tanvir Mahdad, Md Mojibur Rahman Redoy Akanda, Cong Shi, Yan Wang, Nitesh Saxena, and Yingying Chen. 2023. FaceReader: Unobtrusively Mining Vital Signs and Vital Sign Embedded Sensitive Info via AR/VR Motion Sensors. In *Proceedings of the 2023 ACM SIGSAC Conference on Computer and Communications Security* (Copenhagen, Denmark) *(CCS '23)*. Association for Computing Machinery, New York, NY, USA, 446–459. doi:10.1145/3576915.3623102